\documentclass[aps,pra,twocolumn,showpacs,longbibliography]{revtex4-1}

\usepackage{amsmath}
\usepackage{graphicx}
\usepackage{color}
\usepackage{hyperref}

\begin{document}

\title{Lindblad approach to spatio-temporal quantum dynamics\\
of phonon-induced carrier capture processes}

\author{Roberto Rosati}
\affiliation{ Institut f\"ur Festk\"orpertheorie, Universit\"at M\"unster,
Wilhelm-Klemm-Str. 10, 48149 M\"unster, Germany}

\author{Doris E. Reiter}
\affiliation{ Institut f\"ur Festk\"orpertheorie, Universit\"at M\"unster,
Wilhelm-Klemm-Str. 10, 48149 M\"unster, Germany}

\author{Tilmann Kuhn}
\affiliation{ Institut f\"ur Festk\"orpertheorie, Universit\"at M\"unster,
Wilhelm-Klemm-Str. 10, 48149 M\"unster, Germany}

\date{\today}

\begin{abstract}
In view of the ultrashort spatial and temporal scales involved, carrier
capture processes in nanostructures are genuine quantum phenomena. To
describe such processes, methods with different levels of approximations have
been developed. By properly tailoring the Lindblad-based nonlinear  single-particle density
matrix equation provided by an alternative Markov approach, in this work we present a Lindblad superoperator to describe how the phonon-induced carrier capture affects
the spatio-temporal quantum dynamics of a flying wave packet impinging on a
quantum dot. We compare the results with non-Markovian quantum kinetics
calculations and discuss the advantages and drawbacks of the two approaches.
\end{abstract}



\maketitle

\section{Introduction}\label{s-I}

The shrinking of space- and time-scales of modern devices has reached the
threshold where semiclassical approaches like the Boltzmann equation
\cite{b-Jacoboni89} are no longer able to fully catch the genuine quantum
mechanical effects on the electronic transport
\cite{Rossi02b,Axt04a,Pecchia04a}. To model the dynamics in these systems,
several quantum mechanical approaches have been proposed, including
nonequilibrium Green's functions \cite{b-Haug04,b-Bonitz98,b-Datta97}, path
integrals \cite{Makri95,Vagov11,Glaessl11}, surface hopping approaches \cite{Wang15, Wang16} and density matrix based
treatments either on a non-Markovian, quantum kinetic (QK) level
\cite{Schilp94a,Rossi02b,Axt04a} or on a Markovian level
\cite{Fischetti99a,Knezevic08a,Hohenester97a,BiSun99a}.

\begin{figure}[h]
\centering
\includegraphics[width=\linewidth]{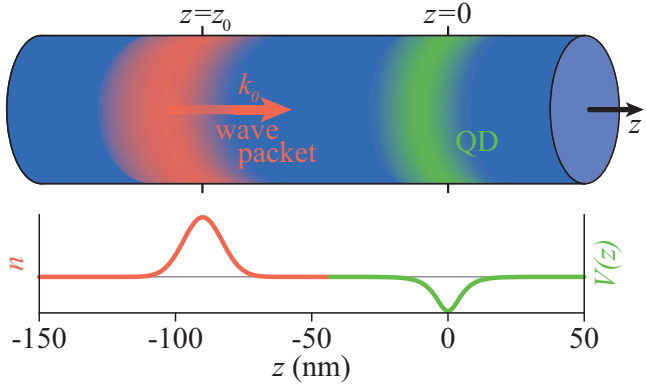}
\caption{
(Color online) Schematic representation of the carrier capture process
from a wave packet (red shapes) traveling in a cylindrical quantum wire
into an embedded QD (green shapes): the upper and lower panels show the process in
three-dimensional space and reduced to the longitudinal direction
$z$, respectively. Here $z_0$ and $k_0$ are the initial central
position and wave vector of the wave packet, respectively (see also Eq.~(\ref{pureState})).
\label{fig:tube}
}
\end{figure}

Carrier capture processes into localized states of a nanostructure
\cite{b-Ferreira15,Glanemann05,Reiter06} are within the most intrinsically
quantum mechanical processes, as can be inferred both from the nanometric
size of its constituents and the subpicosecond time-scale of the interaction.
Furthermore, they involve transitions from extended states in the continuous
part of the spectrum into localized states in the discrete part. In this work we focus on a wave packet traveling in a quantum
wire (QWR) which, when passing by a quantum dot (QD) embedded in the wire,
interacts with the latter by means of electron-phonon scattering, see
Fig.~\ref{fig:tube}. Experimentally, QWRs with an embedded QD have been obtained in different geometries, e.g., by cleaved-edge overgrowth \cite{Wegscheider97}, by growth on a patterned substrate \cite{Lienau00} or by growth in vertical nanowires \cite{Tatebayashi15,Heiss13,Claudon10,Loitsch15}.

The intrinsically spatially inhomogeneous problem induces, in turn, a
nontrivial spatio-temporal dynamics, which could then suggest interesting
applications in electronic-based quantum information processing. For example,
in the concept of flying qubits one could use the shape of a traveling
electronic wave packet to store and transmit information around the
nanodevice \cite{Bertoni00,Ionicioiu01,Feve07}. Potentially, the carrier
capture processes could be able to alter this information in a point which is
strongly localized both in space and time \cite{Glanemann05}; considering the
rise of novel materials able to provide strain-tunable QDs
\cite{Feng12,Manzeli15,Roldan15,Kern16} or dispersionless propagation
\cite{Rosati15,Rosati15b}, this property could make the capture processes one key
ingredient of electronic-based quantum information protocols.

On the other hand, due to the intrinsically local nature of the
carrier-phonon interaction in combination with the nonlocal character of the
continuum states, the spatio-temporal dynamics of carrier capture processes is
extremely demanding to describe on a fully quantum-mechanical level. In
view of their importance for hetero-structure semiconductor lasers, capture
rates obtained from Fermi's Golden Rule (FGR) have been calculated for many
years, first mainly for the capture from bulk into quantum well states
\cite{Brum86,Kuhn89,Preisel94,Reiter09} and then also for the capture into QD states
\cite{Ferreira99,Magnusdottir03,Nielsen04}. The resulting semiclassical
treatments may thus efficiently provide the total captured charge, however
they will typically not be able to properly describe the spatio-temporal
dynamics of the traveling wave packet. In addition, the different effective
dimensionalities of states involved in capture processes give rise to
difficulties already in properly defining the semiclassical equation: in
fact, the scattering rates typically depend on the normalization volume of
the delocalized states,which then has to be fixed by some more or less
rigorous argument. 

This difficulty in describing spatio-temporal dynamics together with
scattering processes is related to the fact that, as long as only occupations
are considered, as is often done in these calculations, the electrons that
occupy continuum states are always completely delocalized. Space dependencies
are then often introduced in a phenomenological, parametric way. However,
spatially inhomogeneous distributions, where the carriers are not completely
delocalized, require superpositions of these states, i.e., off-diagonal
elements in the density matrices defined with respect to these states \cite{Herbst03,Steininger96}. This
is exactly what has been done in QK studies, where it has been shown that the
interaction provides \textit{local} capture dynamics, in contrast to what would be seen within a
diagonal description \cite{Glanemann05,Reiter06,Reiter07a}. A local interaction in space
translates into a finite duration in time, which in turn induces broadened
energy selection rules with respect to the Dirac-delta of conventional FGR
\cite{Glanemann05,Reiter06}. QK treatments provide accurate predictions, but
their computational costs make it complicated to study longer evolutions or
stronger interaction mechanisms, where in addition numerical instabilities, especially on longer time scales,
may appear. They also often prohibit the extension to more complicated
structures, e.g., higher-dimensional systems which cannot be reduced by
spatial symmetries.

Hence it is desirable to have a computationally lighter approach still
capable of describing the spatio-temporal dynamics, thereby being
intermediate between FGR and QK. We look therefore for an approach able to
deal with non-diagonal density matrices through a closed equation of motion:
Conventional Markov approximations
\cite{Fischetti99a,Knezevic08a,Hohenester97a,BiSun99a} could in principle
fulfill these requirements, but the asymmetric shape of their superoperator
may lead to huge instabilities due to their non-positive definite definition
\cite{Rosati14e}. The solution to the instability problem can be given by a
recently introduced Markov approximation \cite{Taj09b}, which in fact is able
to provide a Lindblad-like many-body superoperator and a positive-definite
nonlinear single-particle density matrix equation \cite{Rosati14e}, the
latter being of Lindblad type itself in the low density regime.

In this paper we extend this treatment to carrier-capture processes. We then
compare the resulting Lindblad single-particle (LSP) approach to a full QK
analysis, which can be considered as a benchmark. We show that the LSP
approach catches the essential features of the spatio-temporal dynamics and
further discuss the advantages and drawbacks of the two approaches.

The paper is organized as follows: In Sec.~\ref{sec:th} we present the theory
behind the carrier capture processes, recall the main fundamentals of the QK
treatment and introduce the properly tailored LSP approach. In
Sec.~\ref{sec:res} we study the capture process of a nanometric
flying wave packet impinging on a QD with one (Sec.~\ref{sec:sech}) or two
(Sec.~\ref{sec:quad}) bound states, and finally conclude in
Sec.~\ref{sec:sum}.

\section{Theory}\label{sec:th}

\subsection{Hamiltonians and equations of motion}

The starting point for our description is a proper definition of the
single-particle eigenstates $|\alpha\rangle$ and the associated energy levels
$\epsilon_\alpha$ corresponding to the nanodevice potential profile. We will
reduce our system to be effectively one-dimensional by considering only the
lowest transverse eigenmode of a cylindrical GaAs QWR with 100
nm$^2$ cross section. As a consequence, the state $|\alpha\rangle$
corresponds to the eigenfunction $\psi_\alpha(z) \equiv \langle z \vert
\alpha \rangle$ solving the single-particle Schr\"odinger equation
\begin{equation}\label{eigenEq}
\left[ - \frac{\hbar^2}{2 m^*} \frac{\partial^2}{\partial z^2} + V(z) \right]
\psi_\alpha(z) = \epsilon_\alpha \psi_\alpha(z) \, ,
\end{equation}
with $m^*$ being the effective mass and $V(z)$ the profile of the QD
potential along the longitudinal direction $z$ (see also
Fig.~\ref{fig:tube}).  The single-particle spectrum is discrete for the bound
states ($\epsilon_\alpha < 0$) and continuous for the delocalized states
($\epsilon_\alpha > 0$). The dynamical variable is the single-particle
density operator $\hat \rho$, whose matrix elements $\rho_{\alpha \alpha'}$
are defined as
\begin{equation}\label{rhoAAp}
\rho_{\alpha \alpha'} = \mathrm{Tr}[\hat c_{\alpha'}^\dagger \hat c_\alpha \boldsymbol{\hat \rho}] \, ,
\end{equation}
where $\hat c^\dagger_{\alpha}$ and $\hat c_{\alpha}$ are the creation and
annihilation operators of state $\alpha$, while $\boldsymbol{\hat \rho}$ is
the many-body density matrix containing all the electronic and
phononic degrees of freedom. The dynamics of $\rho$ is given by
\begin{equation}\label{rhoAAp_dyn}
\begin{split}
\frac{d \rho_{\alpha \alpha'}}{dt} =& \mathrm{Tr}\left[\hat c_{\alpha'}^\dagger
\hat c_\alpha \frac{d \boldsymbol{\hat \rho}}{dt} \right] \\
\equiv& \left. \frac{d \rho_{\alpha \alpha'}}{dt} \right|_{\text{free}} + \left.
\frac{d \rho_{\alpha \alpha'}}{dt} \right|_{\text{scat}} \, ,
\end{split}
\end{equation}
where in the last equality we have distinguished between the scattering-free
and electron-phonon induced dynamics, which are defined as
\begin{subequations}
\begin{align}
        \left. \frac{d \rho_{\alpha \alpha'}}{dt} \right|_{\text{free}}&=\frac{1}{\imath \hbar}
        \mathrm{Tr}\left[\hat c_{\alpha'}^\dagger \hat c_\alpha \left[
        \hat H_{\text{e}} + \hat H_{\text{ph}},
        \boldsymbol{\hat \rho }\right] \right], \label{rhoAAp_dyn_free}\\
        \left. \frac{d \rho_{\alpha \alpha'}}{dt} \right|_{\text{scat}}&=\frac{1}{\imath \hbar}
        \mathrm{Tr}\left[\hat c_{\alpha'}^\dagger \hat c_\alpha \left[
        \hat H_{\text{e-ph}},
        \boldsymbol{\hat \rho }\right] \right] \, . \label{rhoAAp_dyn_scat}
\end{align}
\end{subequations}
The Hamiltonians $\hat H_{\text{e}}$ and $\hat H_{\text{ph}}$ appearing in
Eq.~(\ref{rhoAAp_dyn_free}) are the scattering-free electronic and phononic
Hamiltonians,
\begin{align}\label{Hfree}
\hat H_{\text{e}}    =& \sum_\alpha \epsilon_\alpha \hat c^\dagger_{\alpha} \hat c_{\alpha} \quad ,\\
\hat H_{\text{ph}} =& \sum_{\xi}\hat H_{\xi} \equiv \sum_{\xi, \mathbf{q}}
\hbar \omega^\xi_{\mathbf{q}} \hat b^{\xi \dagger}_{\mathbf{q}} \hat b^\xi_{\mathbf{q}} \, ,
\end{align}
with $\xi$ denoting the type of phonon (e.g., optical or acoustic,
longitudinal or transverse, ...), $\mathbf{q}$ the three-dimensional phonon
wave vector and $\hat b^{\xi \dagger}_{\mathbf{q}}$ ($\hat
b^\xi_{\mathbf{q}}$) the creation (annihilation) operator of a phonon of type
$\xi$ and wave vector $\mathbf{q}$.

The scattering-induced dynamics of Eq.~(\ref{rhoAAp_dyn_scat}) is given by
the electron-phonon Hamiltonian $\hat H_{\text{e-ph}}$. In a real-space
representation the interaction Hamiltonian has the form
\begin{equation}\label{H_e-p}
\hat H_{\text{e-ph}} = \int d^3r \hat \Psi^\dagger({\mathbf{r}})
\hat V_{\text{e-ph}}({\mathbf{r}}) \hat \Psi({\mathbf{r}}) \, ,
\end{equation}
where $\hat \Psi^\dagger({\mathbf{r}})$ ($\hat \Psi({\mathbf{r}})$) are the
creation (annihilation) operators for an electron at the position
${\mathbf{r}}$ and $\hat V_{\text{e-ph}}({\mathbf{r}})$ is the phonon-induced
potential acting on the electrons. Its detailed form depends on the phonon
type and the interaction mechanism (e.g., deformation potential,
piezoelectric, polar optical, ...). Equation~(\ref{H_e-p}) clearly shows that
the interaction is local in space, i.e., it connects the annihilation of an
electron at a given position with the creation at the same position.
Inserting the mode representations of electrons and holes, the interaction
Hamiltonian can be written in the more conventional form
\begin{align}\label{H_e-xi}
\hat H_{\text{e-ph}} =& \sum_\xi \hat H_{\text{e}-\xi} \nonumber \\
=& \sum_\xi \sum_{\alpha \alpha', \mathbf{q}} (g^{\xi \mathbf{q}-}_{\alpha \alpha'}
\hat{c}^{\dagger}_\alpha \hat{c}_{\alpha'} \hat b^{\xi}_{\mathbf{q}} +
g^{\xi \mathbf{q}+}_{\alpha \alpha'} \hat c^{\dagger}_{\alpha' }
\hat c_{\alpha} \hat b^{\xi\dagger}_{\mathbf{q}})  \, ,
\end{align}
with $+ (-)$ standing for emission (absorption) of a phonon associated with a
transition from state $\alpha$ to $\alpha'$ ($\alpha'$ to $\alpha$). Hereby
we distinguish between continuous-continuous (CC, both $\alpha$ and $\alpha'$
referring to delocalized states), discrete-discrete (DD, both $\alpha$ and
$\alpha'$ referring to bound states) and continuous-discrete transitions (CD,
with one among $\alpha'$ and $\alpha$ referring to a delocalized and the
other to a bound state). In particular, the latter  are responsible
for the carrier capture processes. In view of the large energetic separation
between the bottom of the delocalized states band and the bound states, the
only phonons able to induce CD transitions are the optical ones. For the interaction matrix element we take the
Fr\"ohlich coupling which is typically the dominant electron-phonon
interaction mechanism in III-V semiconductors. Therefore we consider the longitudinal optical (LO) phonons, whose energy
$E_{LO}\equiv \hbar \omega^{LO}_{\mathbf{q}}$ is in good approximation
$\mathbf{q}$-independent. We will restrict ourselves to
the low temperature limit, in which only (spontaneous) emission processes are
allowed due to the negligible value of the Bose-Einstein distribution
$N_{\hbar \omega^{LO}_{\mathbf{q}}} \equiv N_{E_{LO}} \ll 1$. In principle, also the
Coulomb scattering could be included in the equations, but for low carrier
densities its impact is minor and hence it is neglected here. We furthermore assume that the QWR is of high structural quality, such that on the length scales considered here (a few hundred nm) any disorder effects such as impurity scattering or Anderson localization can be neglected.

\subsection{Description of the spatio-temporal dynamics}

In every CD transition, the coefficients $g^{\xi \mathbf{q}\pm}_{\alpha
\alpha'}$ depend on the overlap between the localized wave function of the
bound state and the product between one electronic delocalized wave function
and one phononic plane wave: as a consequence, the electron-phonon
Hamiltonian of Eq.~(\ref{H_e-xi}) is local. A fully diagonal
approach to the capture processes is unable to catch this locality, even if
the initial state is homogeneous \cite{Reiter07a}: a reduction of the
occupation of a delocalized state due to a capture process will immediately
reduce the electron density in the whole structure, in contrast to the
expectation that initially only regions close to the QD should be
depopulated. Only a non-diagonal treatment is able to correctly model the
locality of the capture processes \cite{Reiter07a}.

The relation between off-diagonal elements and local behavior is a general
feature of density matrix-based descriptions. Given the longitudinal
eigenfunction $\psi_\alpha(z)$ of Eq.~(\ref{eigenEq}), the longitudinal
spatial electron density $n(z)$ is given by
\begin{equation}\label{n(z)}
n = \sum_{\alpha \alpha'} \rho_{\alpha \alpha'} \psi_{\alpha}(z)\psi^*_{\alpha'}(z) \, .
\end{equation}
Above a few meV, the delocalized states $\psi_\alpha(z)$ are essentially
plane waves, thus having a spatially homogeneous square modulus. As a
consequence, a localized wave packet outside the dot can only be described by
including off-diagonal elements of the density matrix (already in the absence
of any scattering mechanism).

Different approaches are possible in order to deal with the huge amount of
degrees of freedom of Eq.~(\ref{rhoAAp_dyn}), which is not closed in the
single-particle density matrix $\rho$ since the terms under the trace also
involve phononic degrees of freedom. QK density matrix approaches rely on a
correlation expansion involving the coupling to an increasing number of
phonon operators. The dynamics of the lower order terms depend on the next
order contributions, giving rise to a (bottom-up) hierarchy which has then to
be truncated at some level. The lowest order contribution
induced by the electron-phonon scattering to the dynamics of $\rho$ depends
on phonon-assisted density matrices
$s^{(\xi)}_{\alpha,\mathbf{q},\alpha'}\equiv \mathrm{Tr}[\hat
c_{\alpha'}^\dagger (\hat b^\xi_{\mathbf{q}} - B^\xi_{\mathbf{q}}) \hat
c_\alpha \boldsymbol{\hat \rho}]$, with $B^\xi_{\mathbf{q}} \equiv
\mathrm{Tr}[\hat b^\xi_{\mathbf{q}} \boldsymbol{\hat \rho}]$ giving the
coherent phonon contributions. A detailed discussion on the QK approach applied to carrier
capture can be found in Refs.~\cite{Glanemann05,Reiter06} and references
therein.

On the other hand, the many-body density matrix appearing in the r.h.s. of
Eq.~(\ref{rhoAAp_dyn}) can be rewritten as an integral over an additional
time $t'$ of $d \boldsymbol{\hat \rho}/dt'$, where the latter may be
rewritten through the Liouville-von Neumann equation as a commutator of
$\boldsymbol{\hat \rho}$. Markov approximations then separate the
time-dependence of $\boldsymbol{\hat \rho}$ into a fast contribution caused
by the scattering-free Hamiltonian, which is taken into account exactly, and
a remaining slow contribution, which is then taken out of the integral
\cite{b-Rossi11}: as a consequence, in contrast to QK approaches Markov
approaches are said to disregard memory effects. However, the procedure of
forgetting the memory of times $t'<t$ is extremely delicate, as the crude one
done in conventional Markov approximations could lead to highly problematic
evolutions \cite{Rosati14e} of the single-particle density matrix $\rho$.

\subsection{LSP approach}\label{sec:LSP}

A recently introduced novel Markov approximation is able to solve these
limitations by providing a Lindblad superoperator at the many-body level
\cite{Taj09b} and a nonlinear but still positive-definite (closed and
non-diagonal) equation at the single-particle one \cite{Rosati14e}. The
latter equation is of Lindblad form in the low-density regime, i.e., when the
generalized Pauli factors may be neglected. The resulting superoperator is
fully microscopic and intrinsically able to consider broadened energy
selection rules by tuning the value of the energy-broadening parameter
$\bar{\epsilon}$ appearing in its coefficients {[see Eq. (\ref{A_LO}) in the appendix]. The details of the full
original Lindblad-based single-particle superoperator can be found in
Ref.~\cite{Rosati14e}, while the here adopted LSP equations of motion are summarized in App. \ref{app:noOsc}. While CC scattering mechanisms
can typically be
described in the so-called \textit{completed collision limit}, i.e.,
$\bar{\epsilon}$ going to zero (as it also happens in FGR), in order to
describe CD transitions the proposed LSP approach has to account for the energy-time uncertainty
through a finite $\bar{\epsilon}$, as will be discussed in more detail in Sec. \ref{sec:broadening}. Although in this work we focus on the low-temperature limit, the proposed equations can easily be extended to finite temperatures. In fact, the original Lindblad approach has already been employed in several room-temperature studies of localized wave packets in various nanosystems and under different scattering mechanisms \cite{Rosati13b,Rosati14a,Rosati14b,Dolcini13a,Rosati15,Rosati15b}, while the carrier capture at finite temperature has been already studied by QK treatments in Refs. \cite{Glanemann04,Reiter07a}.

\section{Capture process from flying wave packets}\label{sec:res}

In this work we will focus on a spatially localized wave packet traveling in
the QWR. As initial condition $\rho^0$ we take a right propagating pure
state with Gaussian distribution in both real and wave vector space,
described by the single particle density matrix
\begin{equation}\label{pureState}
\rho^0_{k_1,k_2} \propto  e^{-\Delta^2_z\left( \frac{k_1+k_2}{2} - k_0 \right)^2}
e^{-\frac{1}{4}\Delta^2_z\left( k_1-k_2\right)^2} e^{-\imath z_0 (k_1-k_2)}  \, ,
\end{equation}
where $\Delta_z$ gives the spatial size of the wave packet, while $z_0$  and
$k_0$ are the initial central position and electronic wave vector,
respectively. The latter can be expressed in terms of the excess energy $E_0$
via $k_0=\sqrt{2 m^* E_0}/\hbar$. We then study the evolution of the density
matrix both within the LSP approach of Eqs.~(\ref{dr_scat})-(\ref{Eb_LO}) and
the QK formalism \cite{Glanemann05,Reiter06}. In all our studies we take
$\Delta_z= 10$~nm and an excess energy resonant with the phonon transition to the lowest bound state,
i.e., $E_0 \approx \epsilon_1 + E_{LO}$ with $E_{LO}=36.4$~meV. The QD will be always
centered at $z=0$, while the initial position of the wave packet will always
be on the l.h.s. of the QD (i.e., $z_0 <0$), in accordance with the fact that
we are considering an only right propagating wave packet (i.e.,
$\rho^0_{k k} \approx 0$ for $k<0$); the magnitude of $|z_0|$ will be
changed in order to study the locality of the carrier capture process.

\subsection{Single bound state in a weakly reflecting potential}\label{sec:sech}

We start the discussion with a QD with only one bound state described by the
potential
\begin{equation}\label{sech}
V(z) = V_0 \, \text{sech}(z/a)
\end{equation}
with $V_0=-30$~meV and $a=4$~nm, which yields $\epsilon_1=-14.4$~meV and
$E_0=22$~meV. Without electron-phonon interaction, such a smoothly varying
potential results in a negligible reflection coefficient, such that an
incident wave packet will be completely transmitted. In order to study the
locality of the carrier capture interaction, we use four different initial
positions, $z_0=\{-90,-110,-130,-150\}$~nm.

\subsubsection{Comparison of LSP and QK approaches}\label{sec:LSPAvsQK}

\begin{figure}[h]
\centering
\includegraphics[width=\linewidth]{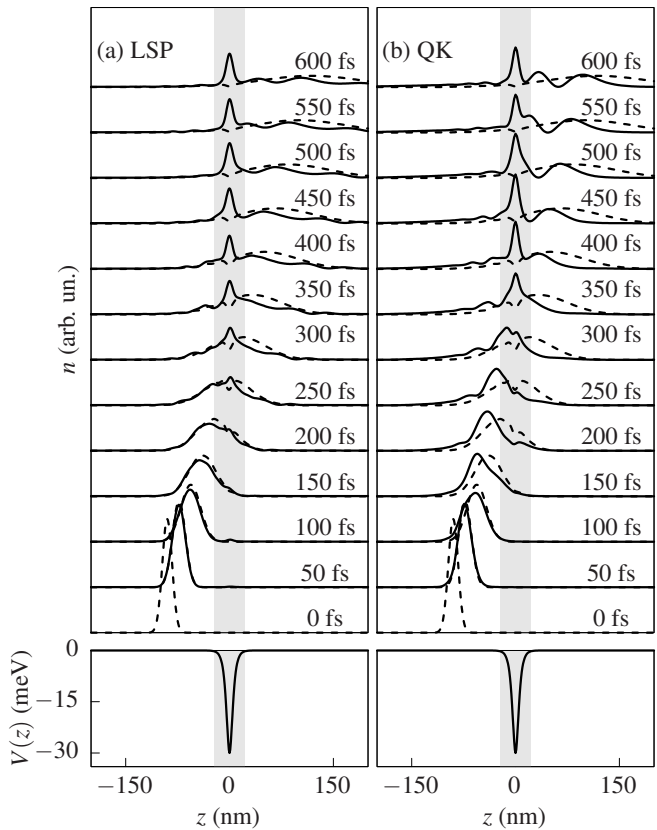}
\caption{Evolution of the spatial charge density $n$ for a traveling
wave packet impinging on the QD (see potential in the bottom panels);
the solid lines show $n$
in the presence of electron-phonon scattering mechanisms as described by the (a) LSP  and (b) QK approach and compared to the scattering-free evolution (dashed lines,
which are identical in the two panels).
\label{fig:sech_n}
}
\end{figure}

In Fig.~\ref{fig:sech_n} we plot the evolution of the spatial charge density
[see Eq.~(\ref{n(z)})] for $z_0=-90$~nm, showing the LSP [panel (a)] and QK
[panel (b)] results (solid lines), while the dashed lines display the free
evolution (which is identical in both cases); the (same) potential profile is
reported in the bottom panels, while the grey shaded background indicates the
QD region as a guide to the eye. In the LSP calculations a broadening of the
DC transitions of $\bar{\epsilon} = 3.5$~meV has been used. The role of this
value will be discussed below in Sec.~\ref{sec:broadening}.

The free evolution allows us to distinguish three phases:
\begin{itemize}
\item[(i)] the travel toward the QD (times $t \lesssim$ 100 fs);
\item[(ii)] the crossing of the QD (100 fs $\lesssim t \lesssim$ 450fs);
\item[(iii)] the moving away from the QD ($t \gtrsim$ 450fs).
\end{itemize}
Note that, even in the absence of scattering, the Gaussian shape is lost
during phase (ii), however it reshapes in phase (iii). Now let us consider
the case with electron-phonon interaction. Due to the capture a density peak
in the QD region builds up during phase (ii) which remains there in phase
(iii). In this last phase (iii), both approaches predict remarkable
differences from the phonon-free case for the transmitted wave packet, in
particular the appearance of spatial ripplings. The shape of the traveling
continuous wave packet is thus altered by the carrier capture process, the
latter happening at a well-defined position (see grey area in
Fig.~\ref{fig:sech_n}) and instant in time (phase (ii)).

While the overall behavior obtained from the two approaches is very similar,
we can also recognize some differences between the respective predictions,
such as, e.g., the height of the spatial rippling of the transmitted wave
packet. In order to better understand the origin of these discrepancies as
well as to better quantify the timescales of the carrier capture, it is
convenient to look at the evolution of the bound state population. In
Fig.~\ref{fig:sech_f1} we plot $f_1 \equiv \rho_{1 1}$ normalized w.r.t. the
preserved total charge $\sum_\alpha \rho_{\alpha \alpha}=\sum_\alpha
\rho^0_{\alpha \alpha}$, as predicted by the LSP and QK approaches (panel
(a) and panel (b), respectively) for four different initial positions $z_0$
of the wave packet. Due to different numerical methods the final populations
(i.e, those after the capture process is over) are slightly different,
although being comparable. Both approaches predict
that the wave packet initially closest (farthest) to the QD is also the first
(last) one to experience the capture mechanism, an additional signature of
the correct description of the local nature of carrier capture processes. The
spatial shift of the initial condition essentially leads to a temporal shift
of the bound state occupation.

In order to be more quantitative on the temporal window in which most of the
carrier capture takes place, in Fig.~\ref{fig:sech_f1_trasl}(a) and (b) we
plot once again the same result of Fig.~\ref{fig:sech_f1} but expressing
each evolution in terms of
\begin{equation}\label{tScal}
\tilde{t}=t+\frac{z_0}{v_{k_0}} \, ,
\end{equation}
with $v_{k_0}=\frac{\hbar k_0}{m^*}$ being the group velocity associated with
$k_0$. $\tilde{t}$ is thus a rescaled time such that the center of each of
the four wave packets, which is roughly moving at a velocity of $v_{k_0}$,
reaches the center of the QD at the same time $\tilde{t} = 0$. In
addition, in Fig.~\ref{fig:sech_f1_trasl} we mark with a background shaded area all
$\tilde{t} \in I_{\tilde{t}}$, with
\begin{equation}\label{It}
I_{\tilde{t}} = \left[ - \frac{\Delta_t}{2}, \frac{\Delta_t}{2} \right] \, ,
\end{equation}
where the value $\Delta_t=\Delta_{\tilde{t}}=235$~fs will be interpreted below when discussing
Fig.~\ref{fig:sech_QMvsSemicl} in terms of the time-evolution of $\left. d
f_1/dt  \right|_{\text{scat}}$ as provided by the LSP approach.

\begin{figure}
\centering
\includegraphics[width=\linewidth]{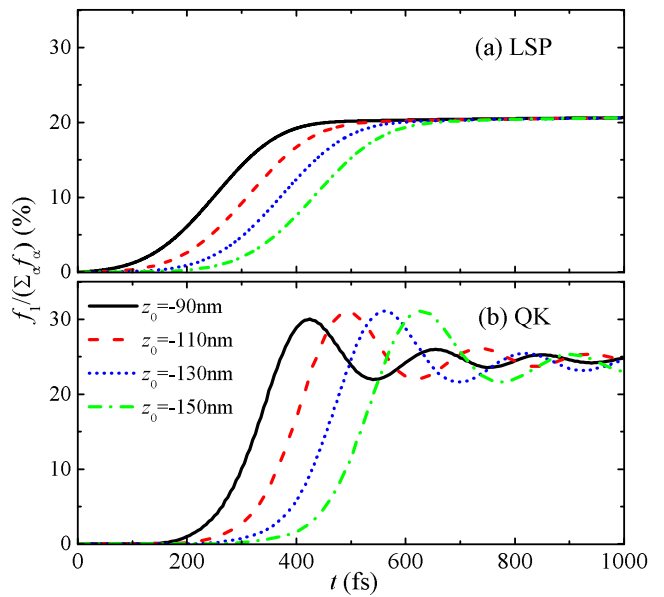}
\caption{
(Color online) Evolution of the normalized bound state population $f_1$ in a
weakly reflecting potential when crossed by wave packets with different initial
positions: $z_0=-90$~nm (solid black line), $z_0=-110$~nm (dashed red line),
$z_0=-130$~nm (dotted blue line) and $z_0=-150$~nm (dashed-dotted green line);
(a) evolution provided by the LSP calculations; (b) evolution obtained within the
QK approach.
\label{fig:sech_f1}
}
\end{figure}

\begin{figure}
\centering
\includegraphics[width=\linewidth]{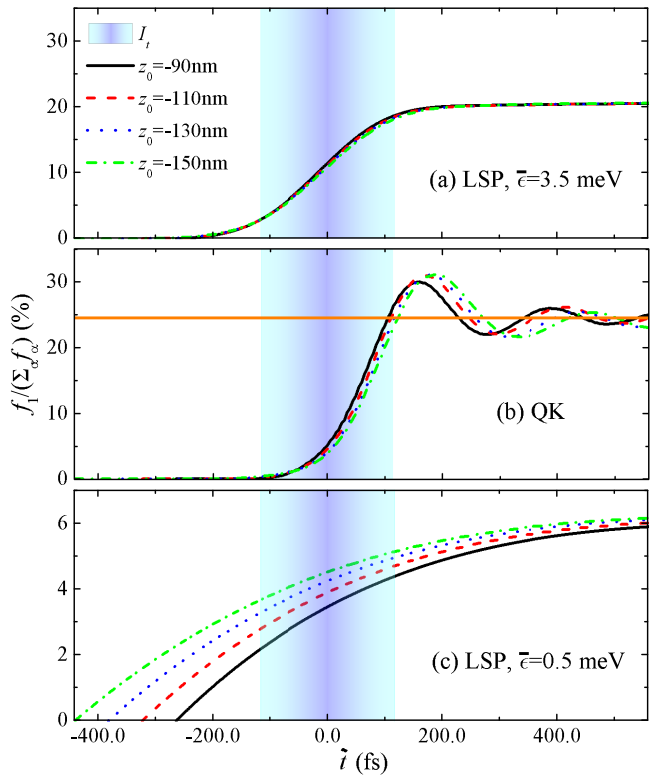}
\caption{(Color online) Same as in Fig.~\ref{fig:sech_f1} but expressing the
evolutions in terms of the scaled time $\tilde{t}$ of Eq.~(\ref{tScal}). (a)
LSP approach; (b) QK approach; (c) LSP approach with a reduced energy broadening
$\bar{\epsilon}=0.5$~meV. The interval
$I_{\tilde{t}}$ defined in Eq.~(\ref{It}) has also been added as a background shaded
area, while an horizontal orange line indicates the stationary value $f_1
\approx 24.5 \%$ in the QK calculations. \label{fig:sech_f1_trasl}}
\end{figure}

Both approaches predict that most of the capture process takes place in the
temporal window defined in Eq.~(\ref{It}). The interval $I_{\tilde{t}}$ can thus be
interpreted as the temporal window in which the carrier capture takes place.
Interestingly, it mostly coincides with phase (ii), which has been defined
through the scattering-free evolution. Going back to the original time for
the initial position $z_0 = -90$~nm adopted in Fig.~\ref{fig:sech_n}, in
fact, this interval goes from $138.5$~fs to $391.5$~fs. However, the behavior
within this interval $I_{\tilde{t}}$ is different in the two approaches.

The evolution provided by the LSP treatment is essentially symmetric around
$\tilde{t} = 0$, where the population is around one half of the final $f_1$:
this  time-symmetric  behavior is not surprising, as the whole Markov
procedure in this approach is based on a temporal symmetrization
\cite{Taj09b}. On the other hand, the evolutions provided by the QK treatment
seem to be slightly retarded: here the population at $\tilde{t} = 0$ is only
around 1/5 of the final value. However, in the second half of $I_{\tilde{t}}$ the
increase of $f_1$ predicted by QK is extremely steep. All the four lines in
Fig.~\ref{fig:sech_f1_trasl}(b) in fact reach the stationary value at
$\tilde{t}=\Delta_t/2$. In phase (ii) also the evolution predicted by the QK
approach is monotonic. The retarded build-up of the bound state occupation
reflects the non-Markovian nature of QK approaches, whose early-stage
dynamics are governed by pseudo-probabilistic rates \cite{b-Rossi11} with
energy selection rules going as $\sin(\omega t)/\omega$ \cite{Schilp94a} and,
as a consequence, they are negligible in the early stages of the interaction.

In phase (iii), i.e., when the carrier capture (shaded area) is over, the QK
treatment predicts a non-monotonic population evolution, while in the LSP
approach the captured density essentially remains constant. The oscillations
in the QK description have been interpreted as phonon-assisted Rabi
oscillations inside the QD \cite{Glanemann05}. The occupation oscillates
periodically between the initial electron state in the continuum and the
correlated state consisting of the electron in the ground state and the
emitted phonon. This behavior is similar to the case of photon-induced Rabi
oscillations of atoms entering a microcavity \cite{Raimond01}. In our case,
the switch on of the electron-phonon interaction is provided by the arrival of the wave packet at the QD, when phonon emission becomes possible. An abrupt switch on thus requires a strongly localized wave packet, which is only possible considering
a broad range of continuum states, whose width in $k$-space is inversely
proportional to the spatial size. The continuum of initial states in turn
generates a continuum of Rabi frequencies with variable detuning. This leads
to the rather strong damping of the oscillations seen in
Figs.~\ref{fig:sech_f1}(b) and \ref{fig:sech_f1_trasl}(b) \cite{Glanemann05}.
As a consequence, these oscillations tend to vanish in few hundreds of fs
after the capture process. Rabi oscillations in optically driven atomic systems are the time-domain counterpart of dressed states, i.e., mixed atom-light states which are split by the light-matter interaction. Analogously, the phonon-assisted Rabi oscillations seen here are the time-domain counterpart of polaronic states in systems with discrete states such as QDs \cite{Verzelen02,Hameau99} or quantum wells in strong magnetic fields \cite{Badalyan88}.

Another difference w.r.t. atoms moving through a
microcavity is the presence of electron-phonon coupling terms which are
diagonal in the electron states. These terms are also not present in
the standard Jaynes-Cummings-like quantum optics models, while in strongly
confined QDs the electron-phonon coupling of this type gives rise to
pure dephasing \cite{Krummheuer02}. Note finally that additional simulations
revealed that these oscillations tend to vanish for increasing size of the
initial wave packet (typically already for widths of the order of 50 nm; not
shown here). More importantly, however, the presence of these oscillations
does not alter the final value of the populations of the bound states.

Such oscillations are not present in the LSP approach. As is shown
analytically in appendix~\ref{app:noOsc}, here the occupation of the bound
state rises monotonically. The absence of Rabi oscillations is not surprising
since, as described above, they rely on the presence of correlated
electron-phonon states, which are not present in the LSP formalism working
completely in the electronic subspace.

The Rabi oscillations in phase (iii) affect the spatial charge density as
well, as they induce the different amplitudes of the transmitted wave-packet
ripplings predicted by the two approaches in Fig.~\ref{fig:sech_n}. The
reason is an enhanced transmission probability each time the electron
oscillates back to the continuum state, which thus results in an additional
peak in the transmitted density seen, e.g., in Fig.~\ref{fig:sech_n} at
$t=600$~fs.

\subsubsection{Non-diagonal vs. diagonal dynamics}\label{sec:QvsSemicl}

In previous works \cite{Reiter07a}, diagonal and non-diagonal QK density
matrix treatments have been compared in terms of the spatial profiles of the
charge density. Here we exploit the proposed LSP treatment in order to
compare diagonal vs. off-diagonal contributions to $\left. d f_1/dt
\right|_{\text{scat}}$: this will allow us to better understand the origin of
the locality and the value of $\Delta_t$, i.e., the duration of phase (ii).

The dynamics of $f_1$
may be rewritten as
\begin{equation}\label{df}
\left. \frac{d f_1}{dt} \right|_{\text{scat}} = \left. \frac{d f_1}{dt}
\right|^{\text{diag.}}_{\text{scat}} + \left. \frac{d f_1}{dt}
\right|^{\text{off-d.}}_{\text{scat}} \quad ,
\end{equation}
where the first (second) term is obtained by considering only the diagonal
(off-diagonal) elements of the density matrices in the r.h.s. of
Eq.~(\ref{dr_scat}).

\begin{figure}
\centering
\includegraphics[width=\linewidth]{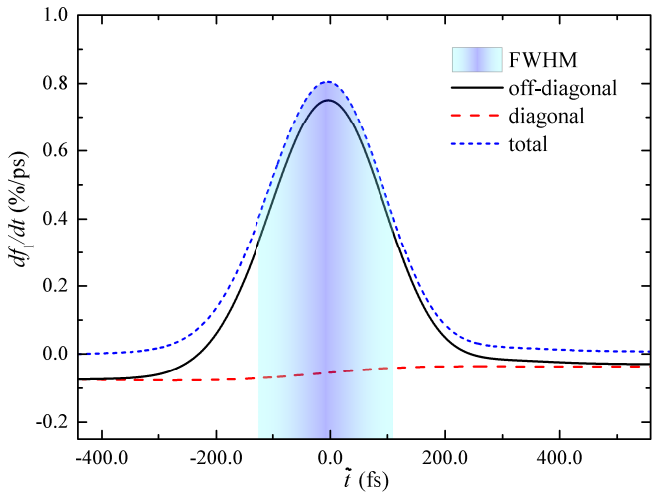}
\caption{
(Color online) Evolution of the off-diagonal (solid black line), diagonal
(long dashed red curve) and total (short-dashed blue line) contributions to the
scattering-induced time-derivative of $f_1$, for the same potential of
Figs.~\ref{fig:sech_f1} and \ref{fig:sech_f1_trasl} and $z_0=-150$~nm; in
order to better show the cancelations between the two terms,
the diagonal contributions are multiplied by $(-1)$.
The background shaded area indicates the
temporal window in which $\left. d f_1/dt \right|_{\text{scat}}$ is bigger than half
its maximum; it has a width of $\Delta_t=235$~fs (FWHM).
\label{fig:sech_QMvsSemicl}
}
\end{figure}

In order to understand why the diagonal contributions are not able to provide the local feature of the capture mechanisms, in
Fig.~\ref{fig:sech_QMvsSemicl} we plot separately the off-diagonal and
diagonal contributions, multiplying the latter by $(-1)$ in order to manifest
the cancelations between the two. As is evident from the long-dashed curve,
the diagonal contribution to the time-derivative of $f_1$ is finite and
positive at any instant, i.e., independent from the fact that at a given
instant the wave packet is in the region of the QD or not. In contrast, the
off-diagonal contributions to $\left. d f_1/dt \right|_{\text{scat}}$ depend
strongly on the position of the wave packet w.r.t. the QD: they provide
negative (positive) values when the wave packet is far from (close to) the
QD. The negative values, occurring when the wave packet is far from the dot,
are exactly the opposite of the diagonal contributions, thus providing a net
cancellation which in turn generates a vanishing derivative of the bound state
population (see short-dashed line). It is thus exactly this different
dependence on the wave-packet position between the diagonal and off-diagonal
contributions in Eq.~(\ref{df}) that gives rise to the local nature of the
capture processes. From a quantitative point of view the
contributions from the off-diagonal elements become much bigger than the
diagonal ones when the wave packet is in the region of the QD, with the
former being about one order of magnitude bigger than the latter around
$\tilde{t} = 0$.

The steepest rise of the bound state occupation occurs when the total
derivative is maximal, which essentially coincides with the time when the
contribution from the off-diagonal terms is biggest. As seen in
Fig.~\ref{fig:sech_QMvsSemicl}, this maximum almost coincides with $\tilde{t}
= 0$, thus inducing the essentially symmetric behavior in time already
discussed after Fig.~\ref{fig:sech_f1_trasl}.

Looking at the total contribution $\left. d f_1/dt\right|_{\text{scat}}$ we
may finally be more quantitative on the duration of phase (ii), i.e., of the
carrier capture process: the temporal length $\Delta_t$ exploited in
Eq.~(\ref{It}) is nothing but the FWHM of the short-dashed line in
Fig.~\ref{fig:sech_QMvsSemicl}. Note that the evaluation of the FWHM of
$\left. d f_1/dt\right|_{\text{scat}}$ is straightforward in the LSP case
because it has only a single relative maximum, while additional arguments or
fittings would be necessary when $f_1$ oscillates in time.

\subsubsection{Role of the energy broadening}\label{sec:broadening}

\begin{figure}
\centering
\includegraphics[width=\linewidth]{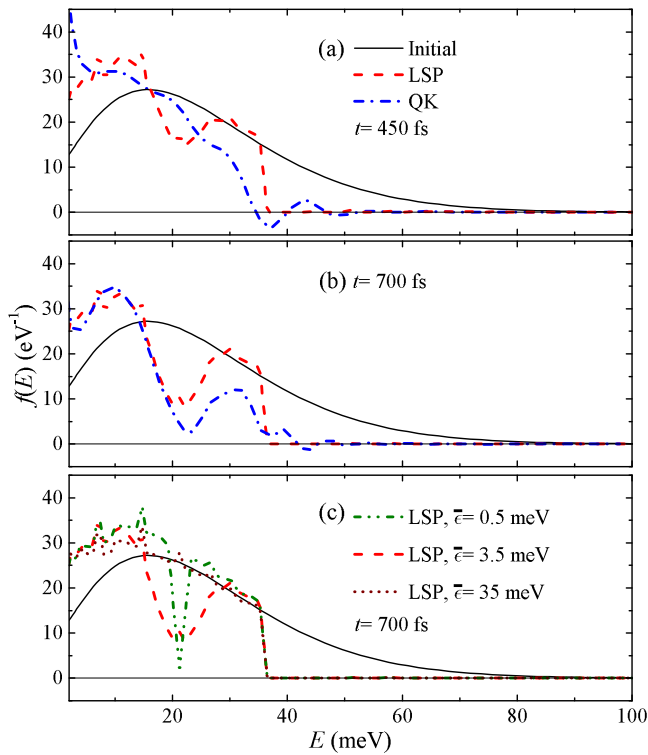}
\caption{(Color online)
Energy distribution in the continuous spectrum for the same potential of
Fig.~\ref{fig:sech_n} with $z_0=-150$~nm at (a) $t=450$~fs and (b),(c)
$t=700$~fs. In (a) and (b) the LSP result with the broadening parameter
$\bar{\epsilon}=3.5$~meV (red dashed line) is compared to the QK result (blue dash-dotted
line). In panel (c) the LSP results for three different broadenings
$\bar{\epsilon}=\{0.5, 3.5, 35\}$~meV are plotted. The solid line in all three
panels shows the initial distribution.
\label{fig:sech_energydistribution}}
\end{figure}

As has been seen in Figs.~\ref{fig:sech_n}--\ref{fig:sech_QMvsSemicl}, The
LSP approach well describes carrier capture except for some phenomena related
to electron-phonon correlations like the phonon-assisted Rabi oscillations,
which are, however, only present for very narrow wave packets. In the
calculations a finite broadening of the CD transitions by
$\bar{\epsilon}=3.5$~meV has been used. In this section we will now analyze
the role of this broadening.

For this purpose, in Figs.~\ref{fig:sech_energydistribution}(a) and (b) we
plot the energy distribution $f(\epsilon_\alpha)=\rho_{\alpha \alpha}/\Delta_{\epsilon_\alpha}$, with $\Delta_{\epsilon_\alpha}$ being the discretization-induced energy uncertainty around state $\alpha$,
 of the continuum states at
two different times $t=450$~fs and $t=700$~fs obtained from the LSP
calculations (dashed lines) and from the QK calculations (dash-dotted lines)
for a wave packet initially located at $z_0 = -150$~nm. The solid line shows
the initial distribution. The first time, $t=450$~fs, corresponds to
$\tilde{t}=8.5$~fs, i.e., essentially to the time when the maximum of the
wave packet is at the QD center, the second time, $t=700$~fs, corresponds to
$\tilde{t}=258.5$~fs, i.e., to a time when the wave packet has passed the QD
region and the capture process is almost completed.

In the LSP calculations we see a sharp cut-off at the LO phonon energy of
$36.4$~meV showing that essentially all carriers above this threshold have
emitted a phonon. Furthermore, at the energy $E_0=\epsilon_1 + E_{LO}=22$~meV
there is a pronounced dip in the distribution caused by the missing carriers
which have been captured. The width of this dip reflects the broadening of of
the transitions by $\bar{\epsilon}=3.5$~meV. Also in the QK calculations the
states above the LO phonon energy are essentially depopulated due to the
strong scattering efficiency. However, the threshold is less pronounced, in
particular at $t=450$~fs when energy-time uncertainty still plays an
important role. At this time there is also no dip due to the captured
electrons visible. In contrast, at $t=700$~fs the threshold behavior builds
up and we observe a clear dip around $E_0$. While in the LSP approach the
broadening has been introduced as a parameter, in the QK approach it is a
result of the quantum dynamics. Here it remains at a finite value because the
capture process has only a limited time window associated with the passage of
the wave packet through the QD region. The broadening seen here motivated us
to the use of the parameter $\bar{\epsilon}=3.5$~meV in the LSP calculations,
because indeed the width of the dips in the distribution are very similar in
both approaches.

As can be seen in Figs.~\ref{fig:sech_energydistribution}(a) and (b), the QK
approach displays some negativities in the electron distributions, in
particular close to the threshold, which are clearly unphysical. The
possibility to obtain such negativities is a consequence of the truncation of
the hierarchy in the QK treatment \cite{Zimmermann94}; they typically vanish
if the truncation is performed at a higher order of the hierarchy
\cite{Krugel06} which, however, in many systems is prohibited by the high
numerical effort. Nevertheless, the magnitude of the negativities should be
kept under control, as bigger values often quickly lead to strong
instabilities. In the present calculations we are still far from such instabilities, as is evident
from the fact that the negativities at $t=700$~fs are even less pronounced
than at $t=450$~fs.

On the other hand, the symmetric structure of the LSP superoperator is able
to guarantee a positive evolution also for arbitrary high interaction
magnitude \cite{Rosati14e}. Interestingly, the proposed approach is Lindblad
already at the lowest approximation level, i.e., uncorrelated phononic and
electronic subspaces, with the former taken as thermalized and the latter
treated within the mean field approximation \cite{Rosati14e}. Thanks to this,
the LSP approach offers a versatile and extremely light alternative to QK for
those situations in which the phonon dynamics or  the electron-phonon
correlations may be disregarded. Note that the proposed approach is
computationally light both in terms of memory requirements, as it does not
require the storage of the phonon-assisted density matrices (necessary on the
other hand in QK) and of simulation time, which for all the results here
reported is of the order of very few minutes for every simulation ran on a
common desktop pc. These computational performances suggest that the LSP approach will be able to describe the carrier capture also in higher dimensional systems. Note that the original Lindblad approach has already been applied to CC transitions between fully three-dimensional states \cite{Dolcini13a}.

To get more insight into the role of the broadening in the LSP calculations,
in Fig.~\ref{fig:sech_energydistribution}(c) we plot the energy distribution
at $t=700$~fs for three different values of $\bar{\epsilon}=\{0.5, 3.5,
35\}$~meV. All other parameters are the same as in
Figs.~~\ref{fig:sech_energydistribution}(a) and (b). When $\bar{\epsilon}=0.5$~meV,
the broadening of the dip has a width of a few meV. For
$\bar{\epsilon}=3.5$~meV the width is about 10~meV. Finally, for
$\bar{\epsilon}=35$~meV the broadening is so large that the capture occurs
from all states leading to a reduced distribution over the whole energy
range. Obviously, the calculations with $\bar{\epsilon}=3.5$~meV exhibit the
best agreement with the QK model.

In addition to the agreement with the QK calculations there is another clear
argument against the use of very narrow, or in the limit delta-like, rates
like in FGR for the CD transitions. This can be seen in
Fig.~\ref{fig:sech_f1_trasl}(c), where we plot the occupation of the bound
state as a function of time in the LSP approach for the same four initial
positions $z_0$ as in Figs.~\ref{fig:sech_f1_trasl}(a),(b), however for the
very small broadening of $\bar{\epsilon}=0.5$~meV. Interestingly, the
occupation starts to grow already from the beginning, long before the wave
packet has reached the QD region. A very small broadening thus also violates
the locality of the capture process, much like the neglect of off-diagonal
terms in the density matrix. In addition, the final value of the bound state
occupation is much less than in the QK case and the LSP case with
$\bar{\epsilon}=3.5$~meV. This can be understood from
Fig.~\ref{fig:sech_energydistribution}(c), which shows that the energy
distribution is completely depopulated at the resonance energy for capture
processes. However, because of the very small broadening there are no more
electrons available to be captured in the bound state, in contrast to the
case of larger broadenings when more electrons are available for the capture
process.

\subsection{Two bound states in a strongly reflecting potential}\label{sec:quad}

In this part we consider the case of a strongly reflecting potential with two
bound states, namely a square well potential with depth of $-25$~meV and a
width of $30$~nm, which has two bound states with energy
$\epsilon_1=-21.4$~meV (thus we take $E_0=15$~meV) and
$\epsilon_2=-11.2$~meV. The initial position of the traveling wave packet is
$z_0=-90$~nm.

\begin{figure}
\centering
\includegraphics[width=\linewidth]{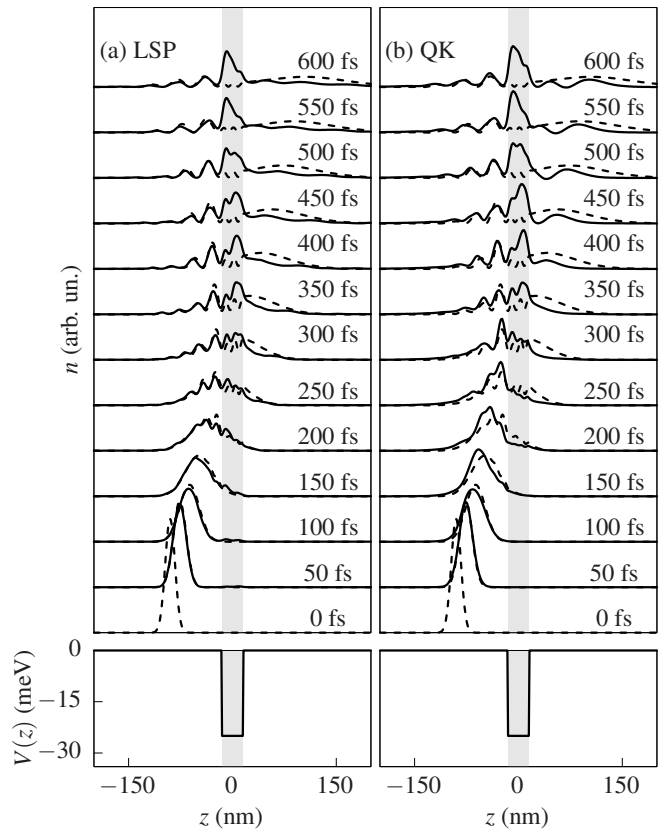}
\caption{
Evolution of the spatial charge density $n$ in the presence of a
strongly reflecting square well potential (shown in the two identical bottom panels)
for a traveling wave packet initially located in $z_0=-90$~nm as predicted by
the LSP [solid lines in panel (a)] and QK approach [solid lines
in panel (b)]. The scattering-free evolution (dashed lines) is identical in both panels.
\label{fig:quad_n}}
\end{figure}

In Fig.~\ref{fig:quad_n} we plot the evolution of the spatial electron density
[see Eq.~(\ref{n(z)})], showing the LSP and QK predictions side by side; the
(same) potential profile is reported in the bottom panels, while a grey
background in correspondence to the QD has been added as a guide to the eye.
The first important characteristic is the shape of the scattering-free
distribution, i.e., the (identical) dashed lines: due to the strongly
reflecting nature of the potential, the wave packet splits between a
transmitted and a reflected component already in the absence of scattering
mechanisms. As a consequence, a fraction of the charge does not enter into
the QD region. The transmitted wave packet has a kinetic energy $\tilde{E}$
of around $20$~meV (as could be inferred from the peak position on the r.h.s.
of the QD in phase (iii)), i.e., it is bigger than $E_0$. As will be clearer
after discussing Figs.~\ref{fig:quad_pol} and \ref{fig:quad_f} below, phase
(ii) is centered around a time $t_0 \approx 275.5 \text{fs}
\approx |z_0|/v_{\tilde{E}}<|z_0|/v_{E_0}$ (see also Sec.~\ref{sec:sech} for
comparison). The reflected component is created at the end of phase (i) and
survives for all the remaining evolution. Already at $t=200$~fs we may notice
the presence of three peaks on the l.h.s. of the QD (centered respectively at
$z=-23$~nm, $-37$~nm and $-50$~nm) moving in left direction with different
velocities  (at  $t=600$ fs they are centered at $z=-41$~nm, $-79$~nm
and $-117$~nm, respectively). Both approaches predict that the reflected
charge is almost unaffected by scattering mechanisms, especially in phase
(iii).

Except for its mean velocity without scattering, the behavior of the
transmitted wave packet is qualitatively similar to the one of
Sec.~\ref{sec:sech}: both approaches predict once again that the carrier
capture induces some spatial ripplings, which are not present in a
scattering-free evolution. The QK approach predicts bigger amplitudes than
the LSP treatment for these oscillations, and the reason lies again in the
phonon-assisted Rabi oscillations. However, from a quantitative point of view
we notice that the height of the transmitted peak is particularly small.
Compared to the case of Sec.~\ref{sec:sech}, this smaller transmitted peak is
due to several reasons, such as more captured charge (as there are more
available bound states) and less charge entering the QD (due to the
reflection).

Concerning the charge within the QD, both approaches predict that it gets
captured in phase (ii), i.e., when the non-reflected component of the wave
packet crosses the QD (see also Figs.~\ref{fig:quad_pol} and \ref{fig:quad_f}
below). In contrast to Sec.~\ref{sec:sech}, here the charge inside the QD
shows an oscillating behavior in time.  The timescales for these oscillations
predicted by the two approaches are very similar: as an example,
Fig.~\ref{fig:quad_n} shows that the left peak becomes the biggest one at
around $500$~fs according to both predictions. The oscillations of the
captured charge density are a genuine quantum mechanical effect induced by
the (scattering-induced) correlations between the bound states.

To study the dynamics of the captured carriers we introduce the charge
density $n_{\text{QD}}$ provided by the bound states
\begin{equation}\label{n_qd}
n_{\text{QD}} = \sum_{i,i'=1}^2 \rho_{i i'} \psi_{i}(z)\psi^*_{i'}(z) \, .
\end{equation}
Except for the transient contributions of the travelling wave packet, $n_{\text{QD}}$ mostly
coincides with the full charge distribution $n$ within the QD. In
order to better understand the oscillatory behavior, we split this density
into two parts according to
\begin{equation}\label{nDot}
n_{\text{QD}} = n_{\text{f}} + n_{\text{p}} \, ,
\end{equation}
where
\begin{equation}\label{nF}
n_{\text{f}} = f_1 |\psi_1(z)|^2 + f_2 |\psi_2(z)|^2
\end{equation}
is the charge given by the populations of the bound states, while
\begin{equation}\label{nP}
n_{\text{p}} = 2 \Re[\rho_{1 2} \psi_1(z) \psi^*_2(z)] 
\end{equation}
($\Re$ denoting the real part)
is the charge contribution given by the polarization $p=\rho_{1 2}$
describing the coherence between the two bound states.

\begin{figure}
\centering
\includegraphics[width=\linewidth]{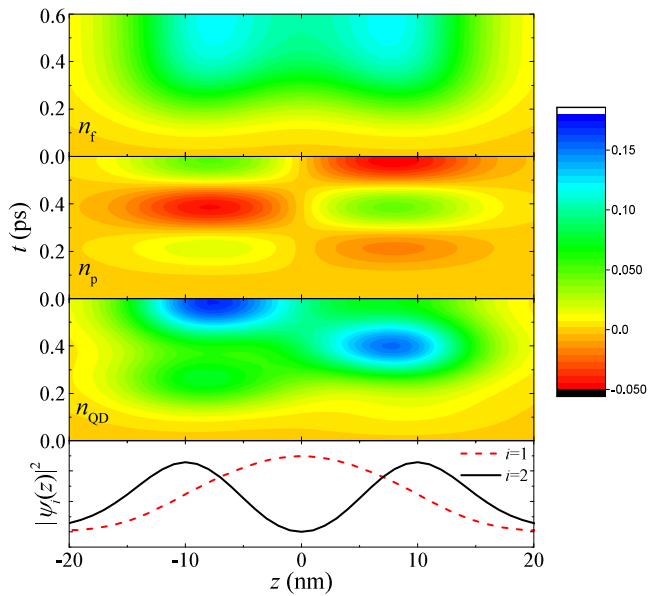}
\caption{
(Color online)
Evolution of $n_{\text{f}}$ (first panel), $n_{\text{p}}$ (second panel) and of
their sum $n_{\text{QD}}$ (third panel) as predicted by the LSP approach for the
same case as in Fig.~\ref{fig:quad_n}. In the fourth panel we report the
spatial profile of $|\psi_i(z)|^2$ for the two bound states, $i=1,2$.
\label{fig:quad_nDot}}
\end{figure}

In Fig.~\ref{fig:quad_nDot} we plot $n_{\text{f}}$, $n_{\text{p}}$ and
$n_{\text{QD}}$ as a function of time $t$ and position $z$ in three different panels. Concerning
the diagonal contribution $n_{\text{f}}$, we see  the appearance of two
peaks close to the two maxima of $|\psi_2(z)|^2$: despite the fact that
$|\psi_1(z)|^2$ has one only peak and that state 1 is more populated than
state 2, $n_{\text{f}}$ has two peaks. Due to the even nature of both
$|\psi_1(z)|^2$ and $|\psi_2(z)|^2$, the spatial distribution
$n_{\text{f}}$ is completely symmetric at any time and stationary after the capture, in contrast to the
electronic density $n$ within the QD region of Fig.~\ref{fig:quad_n}. Since proportional to the product of one even and one odd bound state wavefunction, $n_{\text{p}}$ is on turn spatially anti-symmetric at any time. The temporal dependency of $n_{\text{p}}$ is all included in $\rho_{1 2}$, thus leading to the oscillating behaviour with frequency $(\epsilon_2-\epsilon_1)/\hbar$ provided by the scattering-free evolution of $\rho$. 
Both $n_{\text{f}}$ and $n_{\text{p}}$ have a well defined spatial symmetry, in contrast to the  electronic distribution $n$ within the QD area shown in Fig. \ref{fig:quad_n}.
The discrepancy disappears in
$n_{\text{QD}}$, where, as expected, we recognize the spatio-temporal features of $n$ for small $|z|$.
At around $t$=150 fs we distinguish the appearance of a charge peak around $z\approx -8$~nm;
the latter then moves to the right (following the traveling wave packet) up to around
$400$~fs, where it is centered in $z\approx 8$~nm -- below we will identify 150 fs and 400 fs as the edges of phase
(ii), see Figs.~\ref{fig:quad_pol} and \ref{fig:quad_f}. 
A very similar behavior is seen in the total
captured charge within the QD in Fig.~\ref{fig:quad_n}: in fact, once the
transmitted wave packet leaves the QD, the spatial charge distribution in the
latter is essentially given by $n_{\text{QD}}$. At times slightly bigger
than $400$~fs we are at the beginning of phase (iii) (see also
Figs.~\ref{fig:quad_pol} and \ref{fig:quad_f} below): the traveling wave
packet leaves the QD, while $n_{\text{QD}}$ gets reflected  due to the
interplay between $n_{\text{f}}$ and $n_{\text{p}}$. As an example, we
examine this interplay looking at what happens at $t\approx 600$~fs, when the
peak of $n_{\text{QD}}$ is again within the left half of the QD. This
happens because on the right half of the QD there is a strong cancelation
between $n_{\text{f}}$ and $n_{\text{p}}$; in contrast, in the left
half of the dot the two contributions have the same sign, thus creating a
$n_{\text{QD}}$ higher than the one of $n_{\text{f}}$.

Somehow similarly to what has been seen in Sec.~\ref{sec:QvsSemicl}, we thus
conclude that the local (oscillating) behavior of the carrier capture within
the QD is given by a proper interplay and cancelation between the diagonal
and off-diagonal contributions (in this case to $n_{\text{QD}}$, i.e.
$n_{\text{f}}$ and $n_{\text{p}}$, respectively). However,
$n_{\text{p}}$ affects the spatio-temporal dynamics of the captured
charge, but not its total magnitude: in fact, it is known that the spatial
integral of the off-diagonal contributions to the density matrix vanishes (even in the presence of more bound states),
thus not affecting the total charge brought by $n_{\text{QD}}$,
\begin{equation}\label{nP_int}
\int dz \, n_{\text{p}} = 0 \, \longrightarrow \, \int dz \, n_{\text{QD}} =
\int dz \, n_{\text{f}}  \, .
\end{equation}
As a consequence, the inter-state correlations may shift the charge from
point to point, without however modifying the net amount of total charge.

In order to catch this extremely interesting oscillating behavior, it is 
necessary to predict accurately not only the absolute value of $\rho_{1 2}$,
but also its complex phase. Since both the LSP and QK approach agree in
describing the oscillating evolution of the charge within the QD, it is 
not surprising that the two approaches predict a very similar phase of
$\rho_{1 2}$. This is shown in Fig.~\ref{fig:quad_pol} where, as in Sec.
\ref{sec:sech}, we add a shaded background area indicating phase (ii), which
here is described through the interval $I_{t} =
[t_{0}-\Delta_{t}/2,t_{0}+\Delta_{t}/2]$, where
$t_{0}=275.5$~fs is the time in which $\left. d f_1/dt
\right|_{\text{scat}}$ (evaluated by the LSP approach and not plotted for the
sake of shortness) is maximal, while $\Delta_{t}=276$~fs is the FWHM of
$\left. d f_1/dt  \right|_{\text{scat}}$.

\begin{figure}
\centering
\includegraphics[width=\linewidth]{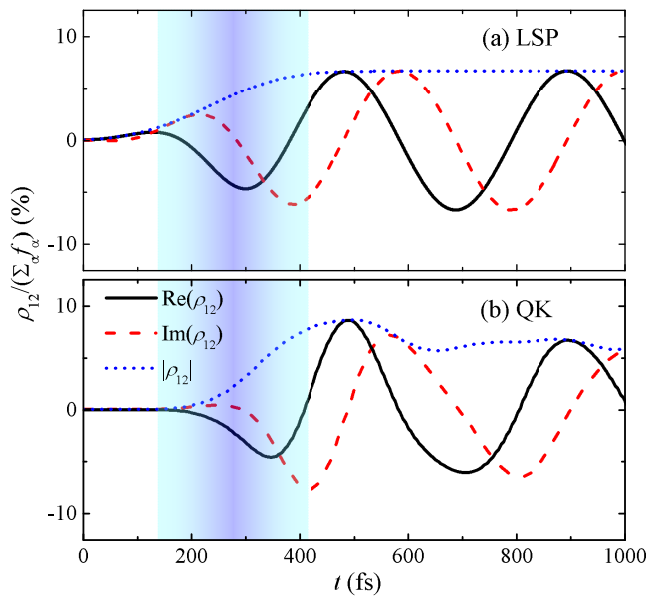}
\caption{
(Color online)
Evolution of the real part (solid black line), imaginary part (dashed
red line) and modulus (dotted blue line) of the polarization $\rho_{12}$ between
the only two bound states for the same case as in Figs. \ref{fig:quad_n} and
\ref{fig:quad_nDot} as predicted by the (a) LSP and (b) QK approach. The background shaded area  indicates phase (ii).
\label{fig:quad_pol}}
\end{figure}

In addition to the matching in terms of complex rotational phase, in
Fig.~\ref{fig:quad_pol} we observe a similar magnitude $|\rho_{1 2}|$, which
is negligible in phase (i), increases in phase (ii) and finally stabilizes in
phase (iii). Interestingly, in phase (ii) the increase of $|\rho_{1 2}|$
predicted by QK is monotonic, similarly to what happens for the population
$f_1$ in Sec.~\ref{sec:sech}.

Figure \ref{fig:quad_pol} then shows that the period of the oscillations agrees
with the Bohr period given by the energy separation between the two states
\cite{Glanemann05}, i.e., $2\pi\hbar/(\epsilon_2 - \epsilon_1)$. Both the
off-diagonal elements shown in Fig.~\ref{fig:quad_pol} and the spatial
oscillations of the captured charge shown in Figs.~\ref{fig:quad_n} and
\ref{fig:quad_nDot} thus demonstrate that after the capture process the
carriers are in a coherent superposition of the bound states. As a
consequence, neither of the two characteristics would be present in a
diagonal density matrix approach.

\begin{figure}
\centering
\includegraphics[width=\linewidth]{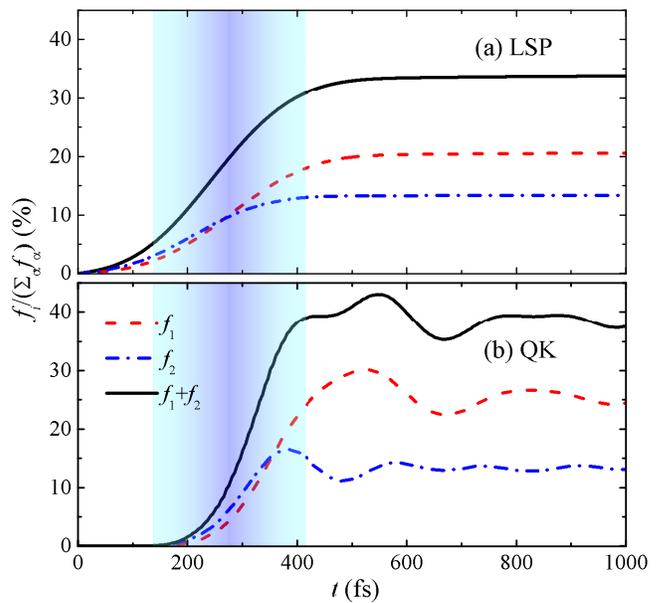}
\caption{
(Color online)
Evolution of the normalized populations of the bound states $f_1$ (dashed red
line) and $f_2$ (dash-dotted blue line) as well as of their sum $f_1+f_2$
(solid black line) as predicted by the (a) LSP and (b) QK approach for the same case as in Figs.~\ref{fig:quad_n}-\ref{fig:quad_pol}. The
background shaded area indicates phase (ii).
\label{fig:quad_f}
}
\end{figure}

Finally, in Fig. \ref{fig:quad_f} we report the evolution of the population
of the bound states. Once again, the evolution of $f_1$ in phase (ii) (i.e.,
$t \in I_t$) predicted by QK is asymmetric w.r.t. $t_0$, although
once again it reaches the stationary values at the end of phase (ii) and the
Rabi oscillations of $f_1$ start only in phase (iii). Both approaches predict
that the evolution of $f_2$ stops earlier than the end of phase (ii). This is
due to the two peaks present in $|\psi_2(z)|^2$: when the wave packet reaches
the second one, it has already given to state 2 most of its energetically
favorable charge while crossing the first one, thus the carrier capture into
state 2 is less efficient for times bigger than $t_0$.
As in Sec. \ref{sec:sech}, also here we find a good agreement between the LSP and QK approach.

\section{Summary and conclusions}\label{sec:sum}

We have presented a single-particle density matrix approach which, due to its
Lindblad-type nature, can describe the spatio-temporal quantum dynamics of
carrier capture processes also in more complicated systems. The LSP approach
is fully non-diagonal, thus able to describe the locality of the capture
process, but also closed in the density matrix and positive-definite, thus
stable and computationally light. The proposed approach is obtained by
tailoring a recently introduced Lindblad-based equation to the specific case
of carrier capture processes, in particular including the broadened energy
selection rules shown by the QK studies.

By comparing the predictions provided by the LSP and QK approaches, we have
identified the role played by various inter-state correlations in carrier
capture processes. Most of the essential features of the spatio-temporal
quantum dynamics of the carrier capture provided by the LSP and QK treatment
are comparable: This is the case for the locality of the interaction (see,
e.g., Fig.~\ref{fig:sech_f1_trasl}), the magnitude of the captured
populations (see Figs.~\ref{fig:sech_f1} and \ref{fig:quad_f}), the magnitude
of inter-bound states correlations (see Fig.~\ref{fig:quad_pol}) and the
spatial evolution (see Figs.~\ref{fig:sech_n} and \ref{fig:quad_n}). In
particular, the very good similarity between the two predicted quantum coherences
between bound states represents a clear fingerprint of the quantum nature of
the approaches, which in turn has a big impact also in terms of the dynamics
of the spatial charge density (see the oscillations of the captured charge in
Figs.~\ref{fig:quad_n} and \ref{fig:quad_nDot}).

In addition to these large similarities, two differences have also been
shown. The first one is a differently distributed efficiency within the
interaction time window: in particular, the evolution predicted by the LSP
approach is symmetric around the time in which the center of the wave packet
reaches the center of the QD, a symmetry stemming from the fully
time-symmetric derivation of the many-body superoperator \cite{Taj09b}. The QK
approach exhibits a retardation related to its non-Markovian nature. The
second difference is the absence, shown also analytically, of the oscillations of the bound states
population after the passage of the wave packet by the QD in the LSP treatment: these
phonon-assisted Rabi oscillations, which tend to vanish in few hundreds of fs
after the capture process, stem from electron-phonon correlations
\cite{Glanemann05}, which are included in the QK treatment but not in the LSP
approach (and more in general in most of the Markov approaches).

In summary, semiclassical treatments are the fastest ones for predicting the
captured charge from homogeneous distributions, while QK approaches offer the
possibility to study the spatio-temporal dynamics including also
electron-phonon correlations and phonon dynamics. The proposed LSP approach
offers an intermediate solution for studying the spatio-temporal quantum
dynamics of the electronic wave packets also in more complicated schemes.
Thanks to its stable and computationally much less-demanding nature, it is
extendible to higher dimensional systems, to longer times and to phenomena
involving different interaction mechanisms. In particular, strain-induced QDs \cite{Feng12,Manzeli15,Roldan15,Kern16}
could give rise to interesting applications where, thanks to the possibility
of a dynamical strain-engineering of the shape of the QD, they could lead to
dynamical modifications of the shape of the wave packet, a key ingredient in
a quantum information protocol based on the electronic charge degree of
freedom.

\begin{acknowledgments}

We thank Daniel Wigger for help with the graphics.

\end{acknowledgments}

\appendix

\section{Lindblad approach to carrier capture}\label{app:noOsc}

In this Appendix we explicitly write the form of the LSP superoperator and
show that it always provides a monotonic rise of the bound states population.
The starting point is the previously reported \cite{Rosati14e} nonlinear
single-particle superoperator, which for the specific case of electron-phonon
scattering and in the  low-density regime, i.e. $|\rho_{\alpha \alpha'}| \ll
1$, may be rewritten omitting the generalized Pauli factors (i.e.
$(\delta_{\alpha \alpha'}-\rho_{\alpha \alpha'})\approx \delta_{\alpha
\alpha'}$) as
\begin{equation}\label{dr_scat}
\left. \frac{d \rho_{\alpha \alpha'}}{dt} \right|_{scat} \!\!\!=
\frac{1}{2}\!\sum_{\bar{\alpha} \bar{\alpha}', s} \!\!\left(
\mathcal{P}^s_{\alpha \alpha' \bar{\alpha} \bar{\alpha}'} \rho_{\bar{\alpha}
\bar{\alpha}'} \!-\! \mathcal{P}^{s*}_{\bar{\alpha}\bar{\alpha}\alpha \bar{\alpha}'}
\rho_{\bar{\alpha}' \alpha'}\right) + \text{H.c.} \, .
\end{equation}
In principle $s=(\xi,\mathbf{q},\pm)$ could label a generic scattering
mechanisms, but in this paper we focus on LO phonon coupling in the $T=$0 K limit in GaAs
QWRs, i.e. $s=(\text{LO},\mathbf{q},+) \equiv \mathbf{q}$. The
generalized scattering rates  $\mathcal{P}^s\equiv \mathcal{P}^\mathbf{q}$
may thus be written as
\begin{equation}\label{Pcal}
\mathcal{P}^\mathbf{q}_{\alpha \alpha' \bar{\alpha} \bar{\alpha}'}=
A^{\mathbf{q}}_{\alpha \bar{\alpha}} A^{\mathbf{q}*}_{\alpha' \bar{\alpha}'} \, ,
\end{equation}
with
\begin{equation}\label{A_LO}
A^{\mathbf{q}}_{\alpha \bar{\alpha}} = \sqrt{\frac{2 \pi}{\hbar}}
g^{\mathbf{q}}_{\alpha \bar{\alpha}}\frac{e^{-\left( \frac{\epsilon_\alpha -
\epsilon_{\bar{\alpha}}+E_{LO}}{2 \bar{\epsilon}} \right)^2}}{(2\pi
\bar{\epsilon}^2)^\frac{1}{4}} \, ,
\end{equation}
where the energetic broadening parameter $\bar{\epsilon}$ is given
by (see discussion in Sec. \ref{sec:th})
\begin{equation}\label{Eb_LO}
\bar{\epsilon} \to \begin{cases}
0 \!\!&, \text{CC, } \min(\alpha,\bar{\alpha}) > n_B \\
\bar{\epsilon} \neq 0 \!\!&,
\text{CD, }  \max(\alpha,\bar{\alpha}) > n_B \geq \min(\alpha,\bar{\alpha}) \, ,\\
\end{cases}
\end{equation}
with $g^{\mathbf{q}}_{\alpha \bar{\alpha}}\equiv g^{LO \mathbf{q}+}_{\alpha
\bar{\alpha}}$ and $n_B$ denoting the number of bound states. Note that in this
work we have considered a QD with energetic separations between localized
states strongly different from $E_{LO}$: as a consequence, we have
disregarded all DD transitions (i.e., scattering mechanisms between bound
states). In general, quasibound states may lead to polaronic contributions
\cite{Magnusdottir02}; however, in this work we focus only on
transitions between clearly delocalized states (e.g., energies bigger than
$10$~meV) and deeply-bound states (e.g., energies smaller than $-10$~meV),
which then allows us to neglect similar polaronic contributions in first
approximation. The set of the Eqs.~(\ref{dr_scat}), (\ref{Pcal}),
(\ref{A_LO}) and (\ref{Eb_LO}) constitutes the LSP approach that we have
used.

We now also prove analytically that the LSP approach always provides
monotonic evolutions of the bound state populations: for this purpose we
focus on the lowest bound state $\vert 1 \rangle$, i.e. $\epsilon_1 <
\epsilon_\alpha$ for all $\alpha \neq 1$, and we rewrite its
scattering-induced time derivative provided by Eq.~(\ref{dr_scat}) as
\begin{equation}\label{df1}
\left. \frac{d f_1}{dt} \right|_{scat} = \frac{1}{2}\sum_{\bar{\alpha} \bar{\alpha}',
\mathbf{q}} \left( \mathcal{P}^{\mathbf{q}}_{1 1 \bar{\alpha} \bar{\alpha}'}
\rho_{\bar{\alpha} \bar{\alpha}'} - \mathcal{P}^{\mathbf{q}*}_{\bar{\alpha}\bar{\alpha}
 1 \bar{\alpha}'} \rho_{\bar{\alpha}' 1}\right) + \text{H.c.} \, .
\end{equation}
Recalling Eqs. (\ref{A_LO}) and (\ref{Pcal}), we note that the generalized
rates entering the second term in the sum of Eq.~(\ref{df1}) are proportional
to Gaussian functions in energy which provide negligible contributions: in
fact,
\begin{equation}\label{df1_out}
\begin{split}
\mathcal{P}^{\mathbf{q}*}_{\bar{\alpha}\bar{\alpha} 1 \bar{\alpha}'}
\propto& \,A^{\mathbf{q}}_{\bar{\alpha} \bar{\alpha}'} g^{\mathbf{q}*}_{\bar{\alpha} 1} e^{-\left( \frac{\epsilon_{\bar{\alpha}} - \epsilon_1 +E_{LO}}{2 \bar{\epsilon}}
\right)^2}\\
 \leq& \,A^{\mathbf{q}}_{\bar{\alpha} \bar{\alpha}'} g^{\mathbf{q}*}_{\bar{\alpha} 1} e^{-\left( \frac{E_{LO}}{2 \bar{\epsilon}} \right)^2} \approx 0 \, ,
\end{split}
\end{equation}
where in the first inequality we used that $\epsilon_1 \leq \epsilon_{\bar{\alpha}}$
by definition, while in the last approximation we used $E_{LO} \gg
\bar{\epsilon}$. Inserting Eq.~(\ref{df1_out}) into Eq.~(\ref{df1}), the latter
reduces to
\begin{equation}\label{df1_proofMonoton}
\begin{split}
\left. \frac{d f_1}{dt} \right|_{scat} =& \sum_{\bar{\alpha} \bar{\alpha}',
\mathbf{q}}  \mathcal{P}^{\mathbf{q}}_{1 1 \bar{\alpha} \bar{\alpha}'}
\rho_{\bar{\alpha} \bar{\alpha}'} \\
=&\sum_{\lambda, \mathbf{q}} \tilde{f}_\lambda |G^{\mathbf{q}}_{1 \lambda}|^2
\geq 0 \, ,
\end{split}
\end{equation}
where $\vert \lambda \rangle$ are the eigenstates of $\hat \rho$ with
eigenvalues $\tilde{f}_\lambda \in \left[ 0,1 \right]$, i.e. $\hat \rho \vert
\lambda \rangle = \tilde{f}_\lambda \vert \lambda \rangle$, and
$G^{\mathbf{q}}_{1 \lambda}=\sum_{\bar{\alpha}} A^{\mathbf{q}}_{1
\bar{\alpha}} U_{\bar{\alpha} \lambda}$, with $U_{\bar{\alpha} \lambda} =
\langle \bar{\alpha} \vert \lambda \rangle$. Although strictly valid for the
deepest bound state, the proof can easily be extended to all the bound states
when the DD transitions are of minor importance, as it is
the case in this work.


\begin{thebibliography}{61}%
\makeatletter
\providecommand \@ifxundefined [1]{%
 \@ifx{#1\undefined}
}%
\providecommand \@ifnum [1]{%
 \ifnum #1\expandafter \@firstoftwo
 \else \expandafter \@secondoftwo
 \fi
}%
\providecommand \@ifx [1]{%
 \ifx #1\expandafter \@firstoftwo
 \else \expandafter \@secondoftwo
 \fi
}%
\providecommand \natexlab [1]{#1}%
\providecommand \enquote  [1]{``#1''}%
\providecommand \bibnamefont  [1]{#1}%
\providecommand \bibfnamefont [1]{#1}%
\providecommand \citenamefont [1]{#1}%
\providecommand \href@noop [0]{\@secondoftwo}%
\providecommand \href [0]{\begingroup \@sanitize@url \@href}%
\providecommand \@href[1]{\@@startlink{#1}\@@href}%
\providecommand \@@href[1]{\endgroup#1\@@endlink}%
\providecommand \@sanitize@url [0]{\catcode `\\12\catcode `\$12\catcode
  `\&12\catcode `\#12\catcode `\^12\catcode `\_12\catcode `\%12\relax}%
\providecommand \@@startlink[1]{}%
\providecommand \@@endlink[0]{}%
\providecommand \url  [0]{\begingroup\@sanitize@url \@url }%
\providecommand \@url [1]{\endgroup\@href {#1}{\urlprefix }}%
\providecommand \urlprefix  [0]{URL }%
\providecommand \Eprint [0]{\href }%
\providecommand \doibase [0]{http://dx.doi.org/}%
\providecommand \selectlanguage [0]{\@gobble}%
\providecommand \bibinfo  [0]{\@secondoftwo}%
\providecommand \bibfield  [0]{\@secondoftwo}%
\providecommand \translation [1]{[#1]}%
\providecommand \BibitemOpen [0]{}%
\providecommand \bibitemStop [0]{}%
\providecommand \bibitemNoStop [0]{.\EOS\space}%
\providecommand \EOS [0]{\spacefactor3000\relax}%
\providecommand \BibitemShut  [1]{\csname bibitem#1\endcsname}%
\let\auto@bib@innerbib\@empty
\bibitem [{\citenamefont {Jacoboni}\ and\ \citenamefont
  {Lugli}(1989)}]{b-Jacoboni89}%
  \BibitemOpen
  \bibfield  {author} {\bibinfo {author} {\bibfnamefont {C.}~\bibnamefont
  {Jacoboni}}\ and\ \bibinfo {author} {\bibfnamefont {P.}~\bibnamefont
  {Lugli}},\ }\href@noop {} {\emph {\bibinfo {title} {The Monte Carlo Method
  for Semiconductor Device Simulation}}}\ (\bibinfo  {publisher} {Springer, Wien},\
  \bibinfo {year} {1989})\BibitemShut {NoStop}%
\bibitem [{\citenamefont {Rossi}\ and\ \citenamefont {Kuhn}(2002)}]{Rossi02b}%
  \BibitemOpen
  \bibfield  {author} {\bibinfo {author} {\bibfnamefont {F.}~\bibnamefont
  {Rossi}}\ and\ \bibinfo {author} {\bibfnamefont {T.}~\bibnamefont {Kuhn}},\
  }\bibfield  {title} {\enquote {\bibinfo {title} {Theory of ultrafast
  phenomena in photoexcited semiconductors},}\ }\href {\doibase
  10.1103/RevModPhys.74.895} {\bibfield  {journal} {\bibinfo  {journal} {Rev.
  Mod. Phys.}\ }\textbf {\bibinfo {volume} {74}},\ \bibinfo {pages} {895--950}
  (\bibinfo {year} {2002})}\BibitemShut {NoStop}%
\bibitem [{\citenamefont {Axt}\ and\ \citenamefont {Kuhn}(2004)}]{Axt04a}%
  \BibitemOpen
  \bibfield  {author} {\bibinfo {author} {\bibfnamefont {V.~M.}\ \bibnamefont
  {Axt}}\ and\ \bibinfo {author} {\bibfnamefont {T.}~\bibnamefont {Kuhn}},\
  }\bibfield  {title} {\enquote {\bibinfo {title} {Femtosecond spectroscopy in
  semiconductors: a key to coherences, correlations and quantum kinetics},}\
  }\href {\doibase 10.1088/0034-4885/67/4/R01} {\bibfield  {journal} {\bibinfo
  {journal} {Rep. Prog. Phys.}\ }\textbf {\bibinfo {volume} {67}},\ \bibinfo
  {pages} {433--512} (\bibinfo {year} {2004})}\BibitemShut {NoStop}%
\bibitem [{\citenamefont {Pecchia}\ and\ \citenamefont
  {Di~Carlo}(2004)}]{Pecchia04a}%
  \BibitemOpen
  \bibfield  {author} {\bibinfo {author} {\bibfnamefont {A.}~\bibnamefont
  {Pecchia}}\ and\ \bibinfo {author} {\bibfnamefont {A.}~\bibnamefont
  {Di~Carlo}},\ }\bibfield  {title} {\enquote {\bibinfo {title} {Atomistic
  theory of transport in organic and inorganic nanostructures},}\ }\href
  {\doibase 10.1088/0034-4885/67/8/R04} {\bibfield  {journal} {\bibinfo
  {journal} {Rep. Prog. Phys.}\ }\textbf {\bibinfo {volume} {67}},\ \bibinfo
  {pages} {1497--1561} (\bibinfo {year} {2004})}\BibitemShut {NoStop}%
\bibitem [{\citenamefont {Haug}\ and\ \citenamefont {Koch}(2004)}]{b-Haug04}%
  \BibitemOpen
  \bibfield  {author} {\bibinfo {author} {\bibfnamefont {H.}~\bibnamefont
  {Haug}}\ and\ \bibinfo {author} {\bibfnamefont {S.W.}\ \bibnamefont {Koch}},\
  }\href@noop {} {\emph {\bibinfo {title} {Quantum Theory of the Optical and
  Electronic Properties of Semiconductors}}}\ (\bibinfo  {publisher} {World
  Scientific, Singapore},\ \bibinfo {year} {2004})\BibitemShut {NoStop}%
\bibitem [{\citenamefont {Bonitz}(1998)}]{b-Bonitz98}%
  \BibitemOpen
  \bibfield  {author} {\bibinfo {author} {\bibfnamefont {M.}~\bibnamefont
  {Bonitz}},\ }\href {http://books.google.it/books?id=B7ON7nhWcDIC} {\emph
  {\bibinfo {title} {Quantum Kinetic Theory}}},\ Teubner-Texte zur Physik\
  (\bibinfo  {publisher} {Teubner, Stuttgart},\ \bibinfo {year} {1998})\BibitemShut
  {NoStop}%
\bibitem [{\citenamefont {Datta}(1997)}]{b-Datta97}%
  \BibitemOpen
  \bibfield  {author} {\bibinfo {author} {\bibfnamefont {S.}~\bibnamefont
  {Datta}},\ }\href {http://books.google.it/books?id=28BC-ofEhvUC} {\emph
  {\bibinfo {title} {Electronic Transport in Mesoscopic Systems}}},\ Cambridge
  Studies in Semiconductor Physics and Microelectronic Engineering\ (\bibinfo
  {publisher} {Cambridge University Press, UK},\ \bibinfo {year}
  {1997})\BibitemShut {NoStop}%
\bibitem [{\citenamefont {Makri}\ and\ \citenamefont
  {Makarov}(1995)}]{Makri95}%
  \BibitemOpen
  \bibfield  {author} {\bibinfo {author} {\bibfnamefont {N.}\ \bibnamefont
  {Makri}}\ and\ \bibinfo {author} {\bibfnamefont {D.~E.}\ \bibnamefont
  {Makarov}},\ }\bibfield  {title} {\enquote {\bibinfo {title} {Tensor
  propagator for iterative quantum time evolution of reduced density matrices.
  I. Theory},}\ }\href@noop {} {\bibfield  {journal} {\bibinfo  {journal} {J. Chem. Phys}\ }\textbf {\bibinfo {volume} {102}},\ \bibinfo
  {pages} {4600} (\bibinfo
  {year} {1995})}\BibitemShut {NoStop}%
\bibitem [{\citenamefont {Vagov}\ \emph {et~al.}(2011)\citenamefont {Vagov},
  \citenamefont {Croitoru}, \citenamefont {Gl\"assl}, \citenamefont {Axt},\
  and\ \citenamefont {Kuhn}}]{Vagov11}%
  \BibitemOpen
  \bibfield  {author} {\bibinfo {author} {\bibfnamefont {A.}~\bibnamefont
  {Vagov}}, \bibinfo {author} {\bibfnamefont {M.~D.}\ \bibnamefont {Croitoru}},
  \bibinfo {author} {\bibfnamefont {M.}~\bibnamefont {Gl\"assl}}, \bibinfo
  {author} {\bibfnamefont {V.~M.}\ \bibnamefont {Axt}}, \ and\ \bibinfo
  {author} {\bibfnamefont {T.}~\bibnamefont {Kuhn}},\ }\bibfield  {title}
  {\enquote {\bibinfo {title} {Real-time path integrals for quantum dots:
  Quantum dissipative dynamics with superohmic environment coupling},}\ }\href
  {\doibase 10.1103/PhysRevB.83.094303} {\bibfield  {journal} {\bibinfo
  {journal} {Phys. Rev. B}\ }\textbf {\bibinfo {volume} {83}},\ \bibinfo
  {pages} {094303} (\bibinfo {year} {2011})}\BibitemShut {NoStop}%
\bibitem [{\citenamefont {Gl\"assl}\ \emph {et~al.}(2011)\citenamefont
  {Gl\"assl}, \citenamefont {Vagov}, \citenamefont {L\"uker}, \citenamefont
  {Reiter}, \citenamefont {Croitoru}, \citenamefont {Machnikowski},
  \citenamefont {Axt},\ and\ \citenamefont {Kuhn}}]{Glaessl11}%
  \BibitemOpen
  \bibfield  {author} {\bibinfo {author} {\bibfnamefont {M.}~\bibnamefont
  {Gl\"assl}}, \bibinfo {author} {\bibfnamefont {A.}~\bibnamefont {Vagov}},
  \bibinfo {author} {\bibfnamefont {S.}~\bibnamefont {L\"uker}}, \bibinfo
  {author} {\bibfnamefont {D.~E.}\ \bibnamefont {Reiter}}, \bibinfo {author}
  {\bibfnamefont {M.~D.}\ \bibnamefont {Croitoru}}, \bibinfo {author}
  {\bibfnamefont {P.}~\bibnamefont {Machnikowski}}, \bibinfo {author}
  {\bibfnamefont {V.~M.}\ \bibnamefont {Axt}}, \ and\ \bibinfo {author}
  {\bibfnamefont {T.}~\bibnamefont {Kuhn}},\ }\bibfield  {title} {\enquote
  {\bibinfo {title} {Long-time dynamics and stationary nonequilibrium of an
  optically driven strongly confined quantum dot coupled to phonons},}\ }\href
  {\doibase 10.1103/PhysRevB.84.195311} {\bibfield  {journal} {\bibinfo
  {journal} {Phys. Rev. B}\ }\textbf {\bibinfo {volume} {84}},\ \bibinfo
  {pages} {195311} (\bibinfo {year} {2011})}\BibitemShut {NoStop}%
\bibitem [{\citenamefont {Wang}\ \emph {et~al.}(2015)\citenamefont {Wang},
  \citenamefont {Prezhdo},\ and\ \citenamefont {Beljonne}}]{Wang15}%
  \BibitemOpen
  \bibfield  {author} {\bibinfo {author} {\bibfnamefont {L.}\ \bibnamefont
  {Wang}}, \bibinfo {author} {\bibfnamefont {O.~V.}\ \bibnamefont {Prezhdo}},
  \ and\ \bibinfo {author} {\bibfnamefont {D.}\ \bibnamefont {Beljonne}},\
  }\bibfield  {title} {\enquote {\bibinfo {title} {Mixed quantum-classical
  dynamics for charge transport in organics},}\ }\href@noop {} {\bibfield
  {journal} {\bibinfo  {journal} {Phys. Chem. Chem. Phys.}\
  }\textbf {\bibinfo {volume} {17}},\ \bibinfo {pages} {12395--12406} (\bibinfo
  {year} {2015})}\BibitemShut {NoStop}%
\bibitem [{\citenamefont {Wang}\ \emph {et~al.}(2016)\citenamefont {Wang},
  \citenamefont {Akimov},\ and\ \citenamefont {Prezhdo}}]{Wang16}%
  \BibitemOpen
  \bibfield  {author} {\bibinfo {author} {\bibfnamefont {L.}\ \bibnamefont
  {Wang}}, \bibinfo {author} {\bibfnamefont {A.}\ \bibnamefont {Akimov}}, \
  and\ \bibinfo {author} {\bibfnamefont {O.~V.}\ \bibnamefont {Prezhdo}},\
  }\bibfield  {title} {\enquote {\bibinfo {title} {Recent progress in surface
  hopping: 2011-2015},}\ }\href {\doibase 10.1021/acs.jpclett.6b00710}
  {\bibfield  {journal} {\bibinfo  {journal} {J. Phys. Chem. Lett.}\ }\textbf {\bibinfo {volume} {7}},\ \bibinfo {pages} {2100--2112}
  (\bibinfo {year} {2016})}\BibitemShut {NoStop}%
\bibitem [{\citenamefont {Schilp}\ \emph {et~al.}(1994)\citenamefont {Schilp},
  \citenamefont {Kuhn},\ and\ \citenamefont {Mahler}}]{Schilp94a}%
  \BibitemOpen
  \bibfield  {author} {\bibinfo {author} {\bibfnamefont {J.}~\bibnamefont
  {Schilp}}, \bibinfo {author} {\bibfnamefont {T.}~\bibnamefont {Kuhn}}, \ and\
  \bibinfo {author} {\bibfnamefont {G.}~\bibnamefont {Mahler}},\ }\bibfield
  {title} {\enquote {\bibinfo {title} {Electron-phonon quantum kinetics in
  pulse-excited semiconductors: Memory and renormalization effects},}\ }\href
  {\doibase 10.1103/PhysRevB.50.5435} {\bibfield  {journal} {\bibinfo
  {journal} {Phys. Rev. B}\ }\textbf {\bibinfo {volume} {50}},\ \bibinfo
  {pages} {5435--5447} (\bibinfo {year} {1994})}\BibitemShut {NoStop}%
\bibitem [{\citenamefont {Fischetti}(1999)}]{Fischetti99a}%
  \BibitemOpen
  \bibfield  {author} {\bibinfo {author} {\bibfnamefont {M.~V.}\ \bibnamefont
  {Fischetti}},\ }\bibfield  {title} {\enquote {\bibinfo {title}
  {Master-equation approach to the study of electronic transport in small
  semiconductor devices},}\ }\href {\doibase 10.1103/PhysRevB.59.4901}
  {\bibfield  {journal} {\bibinfo  {journal} {Phys. Rev. B}\ }\textbf {\bibinfo
  {volume} {59}},\ \bibinfo {pages} {4901--4917} (\bibinfo {year}
  {1999})}\BibitemShut {NoStop}%
\bibitem [{\citenamefont {Knezevic}(2008)}]{Knezevic08a}%
  \BibitemOpen
  \bibfield  {author} {\bibinfo {author} {\bibfnamefont {I.}~\bibnamefont
  {Knezevic}},\ }\bibfield  {title} {\enquote {\bibinfo {title} {Decoherence
  due to contacts in ballistic nanostructures},}\ }\href {\doibase
  10.1103/PhysRevB.77.125301} {\bibfield  {journal} {\bibinfo  {journal} {Phys.
  Rev. B}\ }\textbf {\bibinfo {volume} {77}},\ \bibinfo {pages} {125301}
  (\bibinfo {year} {2008})}\BibitemShut {NoStop}%
\bibitem [{\citenamefont {Hohenester}\ and\ \citenamefont
  {P\"otz}(1997)}]{Hohenester97a}%
  \BibitemOpen
  \bibfield  {author} {\bibinfo {author} {\bibfnamefont {U.}~\bibnamefont
  {Hohenester}}\ and\ \bibinfo {author} {\bibfnamefont {W.}~\bibnamefont
  {P\"otz}},\ }\bibfield  {title} {\enquote {\bibinfo {title} {Density-matrix
  approach to nonequilibrium free-carrier screening in semiconductors},}\
  }\href {\doibase 10.1103/PhysRevB.56.13177} {\bibfield  {journal} {\bibinfo
  {journal} {Phys. Rev. B}\ }\textbf {\bibinfo {volume} {56}},\ \bibinfo
  {pages} {13177--13189} (\bibinfo {year} {1997})}\BibitemShut {NoStop}%
\bibitem [{\citenamefont {Bi~Sun}\ and\ \citenamefont
  {Milburn}(1999)}]{BiSun99a}%
  \BibitemOpen
  \bibfield  {author} {\bibinfo {author} {\bibfnamefont {He}~\bibnamefont
  {Bi~Sun}}\ and\ \bibinfo {author} {\bibfnamefont {G.~J.}\ \bibnamefont
  {Milburn}},\ }\bibfield  {title} {\enquote {\bibinfo {title} {Quantum
  open-systems approach to current noise in resonant tunneling junctions},}\
  }\href {\doibase 10.1103/PhysRevB.59.10748} {\bibfield  {journal} {\bibinfo
  {journal} {Phys. Rev. B}\ }\textbf {\bibinfo {volume} {59}},\ \bibinfo
  {pages} {10748--10756} (\bibinfo {year} {1999})}\BibitemShut {NoStop}%
\bibitem [{\citenamefont {Ferreira}\ and\ \citenamefont
  {Bastard}(2015)}]{b-Ferreira15}%
  \BibitemOpen
  \bibfield  {author} {\bibinfo {author} {\bibfnamefont {R.}~\bibnamefont
  {Ferreira}}\ and\ \bibinfo {author} {\bibfnamefont {G.}~\bibnamefont
  {Bastard}},\ }\href@noop {} {\emph {\bibinfo
  {title} {Capture and Relaxation in Self-Assembled Semiconductor Quantum
  Dots}}},\ (\bibinfo  {publisher} {Morgan and Claypool
  Publishers},\ \bibinfo {year} {2015})\BibitemShut {NoStop}%
\bibitem [{\citenamefont {Glanemann}\ \emph {et~al.}(2005)\citenamefont
  {Glanemann}, \citenamefont {Axt},\ and\ \citenamefont {Kuhn}}]{Glanemann05}%
  \BibitemOpen
  \bibfield  {author} {\bibinfo {author} {\bibfnamefont {M.}~\bibnamefont
  {Glanemann}}, \bibinfo {author} {\bibfnamefont {V.~M.}\ \bibnamefont {Axt}},
  \ and\ \bibinfo {author} {\bibfnamefont {T.}~\bibnamefont {Kuhn}},\
  }\bibfield  {title} {\enquote {\bibinfo {title} {Transport of a wave packet
  through nanostructures: Quantum kinetics of carrier capture processes},}\
  }\href {\doibase 10.1103/PhysRevB.72.045354} {\bibfield  {journal} {\bibinfo
  {journal} {Phys. Rev. B}\ }\textbf {\bibinfo {volume} {72}},\ \bibinfo
  {pages} {045354} (\bibinfo {year} {2005})}\BibitemShut {NoStop}%
\bibitem [{\citenamefont {Reiter}\ \emph {et~al.}(2006)\citenamefont {Reiter},
  \citenamefont {Glanemann}, \citenamefont {Axt},\ and\ \citenamefont
  {Kuhn}}]{Reiter06}%
  \BibitemOpen
  \bibfield  {author} {\bibinfo {author} {\bibfnamefont {D.}~\bibnamefont
  {Reiter}}, \bibinfo {author} {\bibfnamefont {M.}~\bibnamefont {Glanemann}},
  \bibinfo {author} {\bibfnamefont {V.~M.}\ \bibnamefont {Axt}}, \ and\
  \bibinfo {author} {\bibfnamefont {T.}~\bibnamefont {Kuhn}},\ }\bibfield
  {title} {\enquote {\bibinfo {title} {Controlling the capture dynamics of
  traveling wave packets into a quantum dot},}\ }\href {\doibase
  10.1103/PhysRevB.73.125334} {\bibfield  {journal} {\bibinfo  {journal} {Phys.
  Rev. B}\ }\textbf {\bibinfo {volume} {73}},\ \bibinfo {pages} {125334}
  (\bibinfo {year} {2006})}\BibitemShut {NoStop}%
\bibitem [{\citenamefont {Wegscheider}\ \emph {et~al.}(1997)\citenamefont
  {Wegscheider}, \citenamefont {Schedelbeck}, \citenamefont {Abstreiter},
  \citenamefont {Rother},\ and\ \citenamefont {Bichler}}]{Wegscheider97}%
  \BibitemOpen
  \bibfield  {author} {\bibinfo {author} {\bibfnamefont {W.}~\bibnamefont
  {Wegscheider}}, \bibinfo {author} {\bibfnamefont {G.}~\bibnamefont
  {Schedelbeck}}, \bibinfo {author} {\bibfnamefont {G.}~\bibnamefont
  {Abstreiter}}, \bibinfo {author} {\bibfnamefont {M.}~\bibnamefont {Rother}},
  \ and\ \bibinfo {author} {\bibfnamefont {M.}~\bibnamefont {Bichler}},\
  }\bibfield  {title} {\enquote {\bibinfo {title} {Atomically precise
  GaAs/AlGaAs quantum dots fabricated by twofold cleaved edge overgrowth},}\
  }\href {\doibase 10.1103/PhysRevLett.79.1917} {\bibfield  {journal} {\bibinfo
   {journal} {Phys. Rev. Lett.}\ }\textbf {\bibinfo {volume} {79}},\ \bibinfo
  {pages} {1917--1920} (\bibinfo {year} {1997})}\BibitemShut {NoStop}%
\bibitem [{\citenamefont {Lienau}\ \emph {et~al.}(2000)\citenamefont {Lienau},
  \citenamefont {Emiliani}, \citenamefont {Guenther}, \citenamefont {Intonti},
  \citenamefont {Elsaesser}, \citenamefont {Nötzel},\ and\ \citenamefont
  {Ploog}}]{Lienau00}%
  \BibitemOpen
  \bibfield  {author} {\bibinfo {author} {\bibfnamefont {Ch.}\ \bibnamefont
  {Lienau}}, \bibinfo {author} {\bibfnamefont {V.}~\bibnamefont {Emiliani}},
  \bibinfo {author} {\bibfnamefont {T.}~\bibnamefont {Guenther}}, \bibinfo
  {author} {\bibfnamefont {F.}~\bibnamefont {Intonti}}, \bibinfo {author}
  {\bibfnamefont {T.}~\bibnamefont {Elsaesser}}, \bibinfo {author}
  {\bibfnamefont {R.}~\bibnamefont {N\"otzel}}, \ and\ \bibinfo {author}
  {\bibfnamefont {K.H.}\ \bibnamefont {Ploog}},\ }\bibfield  {title} {\enquote
  {\bibinfo {title} {Near field optical spectroscopy of confined excitons},}\
  }\href {\doibase 10.1002/1521-396X(200003)178:1<471::AID-PSSA471>3.0.CO;2-Q}
  {\bibfield  {journal} {\bibinfo  {journal} {Phys. Status Solidi A}\
  }\textbf {\bibinfo {volume} {178}},\ \bibinfo {pages} {471--479} (\bibinfo
  {year} {2000})}\BibitemShut {NoStop}%
\bibitem [{\citenamefont {Tatebayashi}\ \emph {et~al.}(2015)\citenamefont
  {Tatebayashi}, \citenamefont {Kako}, \citenamefont {Ho}, \citenamefont {Ota},
  \citenamefont {Iwamoto},\ and\ \citenamefont {Arakawa}}]{Tatebayashi15}%
  \BibitemOpen
  \bibfield  {author} {\bibinfo {author} {\bibfnamefont {J.}\ \bibnamefont
  {Tatebayashi}}, \bibinfo {author} {\bibfnamefont {S.}\ \bibnamefont
  {Kako}}, \bibinfo {author} {\bibfnamefont {J.}\ \bibnamefont {Ho}},
  \bibinfo {author} {\bibfnamefont {Y.}\ \bibnamefont {Ota}}, \bibinfo
  {author} {\bibfnamefont {S.}\ \bibnamefont {Iwamoto}}, \ and\ \bibinfo
  {author} {\bibfnamefont {Y.}\ \bibnamefont {Arakawa}},\ }\bibfield
  {title} {\enquote {\bibinfo {title} {Room-temperature lasing in a single
  nanowire with quantum dots},}\ }\href@noop {} {\bibfield  {journal} {\bibinfo
   {journal} {Nat. Photon.}\ }\textbf {\bibinfo {volume} {9}},\ \bibinfo
  {pages} {501--505} (\bibinfo {year} {2015})}\BibitemShut {NoStop}%
\bibitem [{\citenamefont {Heiss}\ \emph {et~al.}(2013)\citenamefont {Heiss},
  \citenamefont {Fontana}, \citenamefont {Gustafsson}, \citenamefont
  {W{\"u}st}, \citenamefont {Magen}, \citenamefont {O’regan}, \citenamefont
  {Luo}, \citenamefont {Ketterer}, \citenamefont {Conesa-Boj}, \citenamefont
  {Kuhlmann} \emph {et~al.}}]{Heiss13}%
  \BibitemOpen
  \bibfield  {author} {\bibinfo {author} {\bibfnamefont {M.}~\bibnamefont
  {Heiss}}, \bibinfo {author} {\bibfnamefont {Y.}~\bibnamefont {Fontana}},
  \bibinfo {author} {\bibfnamefont {A.}\ \bibnamefont {Gustafsson}},
  \bibinfo {author} {\bibfnamefont {G.}~\bibnamefont {W{\"u}st}}, \bibinfo
  {author} {\bibfnamefont {C.}~\bibnamefont {Magen}}, \bibinfo {author}
  {\bibfnamefont {D.~D.}~\bibnamefont {O'Regan}}, \bibinfo {author}
  {\bibfnamefont {J.~W.}~\bibnamefont {Luo}}, \bibinfo {author} {\bibfnamefont
  {B.}~\bibnamefont {Ketterer}}, \bibinfo {author} {\bibfnamefont
  {S.}~\bibnamefont {Conesa-Boj}}, \bibinfo {author} {\bibfnamefont
  {A.~V.}~\bibnamefont {Kuhlmann}}, \bibinfo {author} {\bibfnamefont
  {J.}~\bibnamefont {Houel}}, \bibinfo {author} {\bibfnamefont
  {E.}~\bibnamefont {Russo-Averchi}},  \bibinfo {author} {\bibfnamefont
  {J.~R.}~\bibnamefont {Morante}}, \bibinfo {author} {\bibfnamefont
  {M.}~\bibnamefont {Cantoni}}, \bibinfo {author} {\bibfnamefont
  {N.}~\bibnamefont {Marzari}}, \bibinfo {author} {\bibfnamefont
  {J.}~\bibnamefont {Arbiol}}, \bibinfo {author} {\bibfnamefont
  {A.}~\bibnamefont {Zunger}}, \bibinfo {author} {\bibfnamefont
  {R.~J.}~\bibnamefont {Warburton}}, \ and\ \bibinfo {author} {\bibfnamefont
  {A.}~\bibnamefont {Fontcubert~i~Morral}},\ }\bibfield  {title}
  {\enquote {\bibinfo {title} {Self-assembled quantum dots in a nanowire system
  for quantum photonics},}\ }\href@noop {} {\bibfield  {journal} {\bibinfo
  {journal} {Nat. Mater.}\ }\textbf {\bibinfo {volume} {12}},\ \bibinfo
  {pages} {439--444} (\bibinfo {year} {2013})}\BibitemShut {NoStop}%
\bibitem [{\citenamefont {Claudon}\ \emph {et~al.}(2010)\citenamefont
  {Claudon}, \citenamefont {Bleuse}, \citenamefont {Malik}, \citenamefont
  {Bazin}, \citenamefont {Jaffrennou}, \citenamefont {Gregersen}, \citenamefont
  {Sauvan}, \citenamefont {Lalanne},\ and\ \citenamefont
  {G{\'e}rard}}]{Claudon10}%
  \BibitemOpen
  \bibfield  {author} {\bibinfo {author} {\bibfnamefont {J.}\ \bibnamefont
  {Claudon}}, \bibinfo {author} {\bibfnamefont {J.}\ \bibnamefont
  {Bleuse}}, \bibinfo {author} {\bibfnamefont {N.~S.}\ \bibnamefont
  {Malik}}, \bibinfo {author} {\bibfnamefont {M.}\ \bibnamefont {Bazin}},
  \bibinfo {author} {\bibfnamefont {P.}\ \bibnamefont {Jaffrennou}},
  \bibinfo {author} {\bibfnamefont {N.}\ \bibnamefont {Gregersen}}, \bibinfo
  {author} {\bibfnamefont {C.}\ \bibnamefont {Sauvan}}, \bibinfo
  {author} {\bibfnamefont {P.}\ \bibnamefont {Lalanne}}, \ and\ \bibinfo
  {author} {\bibfnamefont {J.-M.}\ \bibnamefont {G{\'e}rard}},\
  }\bibfield  {title} {\enquote {\bibinfo {title} {A highly efficient
  single-photon source based on a quantum dot in a photonic nanowire},}\
  }\href@noop {} {\bibfield  {journal} {\bibinfo  {journal} {Nat. Photon.}\
  }\textbf {\bibinfo {volume} {4}},\ \bibinfo {pages} {174--177} (\bibinfo
  {year} {2010})}\BibitemShut {NoStop}%
\bibitem [{\citenamefont {Loitsch}\ \emph {et~al.}(2015)\citenamefont
  {Loitsch}, \citenamefont {Winnerl}, \citenamefont {Grimaldi}, \citenamefont
  {Wierzbowski}, \citenamefont {Rudolph}, \citenamefont {Morkötter},
  \citenamefont {Döblinger}, \citenamefont {Abstreiter}, \citenamefont
  {Koblmüller},\ and\ \citenamefont {Finley}}]{Loitsch15}%
  \BibitemOpen
  \bibfield  {author} {\bibinfo {author} {\bibfnamefont {B.}\
  \bibnamefont {Loitsch}}, \bibinfo {author} {\bibfnamefont {J.}\
  \bibnamefont {Winnerl}}, \bibinfo {author} {\bibfnamefont {G.}\
  \bibnamefont {Grimaldi}}, \bibinfo {author} {\bibfnamefont {J.}\
  \bibnamefont {Wierzbowski}}, \bibinfo {author} {\bibfnamefont {D.}\
  \bibnamefont {Rudolph}}, \bibinfo {author} {\bibfnamefont {S.}\
  \bibnamefont {Mork\"otter}}, \bibinfo {author} {\bibfnamefont {M.}\
  \bibnamefont {D\"oblinger}}, \bibinfo {author} {\bibfnamefont {G.}\
  \bibnamefont {Abstreiter}}, \bibinfo {author} {\bibfnamefont {G.}\
  \bibnamefont {Koblm\"uller}}, \ and\ \bibinfo {author} {\bibfnamefont
  {J.~J.}\ \bibnamefont {Finley}},\ }\bibfield  {title} {\enquote
  {\bibinfo {title} {Crystal phase quantum dots in the ultrathin core of
  GaAs-AlGaAs core-shell nanowires},}\ }\href {\doibase
  10.1021/acs.nanolett.5b03273} {\bibfield  {journal} {\bibinfo  {journal}
  {Nano Lett.}\ }\textbf {\bibinfo {volume} {15}},\ \bibinfo {pages}
  {7544--7551} (\bibinfo {year} {2015})} \BibitemShut {NoStop}%
\bibitem [{\citenamefont {Bertoni}\ \emph {et~al.}(2000)\citenamefont
  {Bertoni}, \citenamefont {Bordone}, \citenamefont {Brunetti}, \citenamefont
  {Jacoboni},\ and\ \citenamefont {Reggiani}}]{Bertoni00}%
  \BibitemOpen
  \bibfield  {author} {\bibinfo {author} {\bibfnamefont {A.}~\bibnamefont
  {Bertoni}}, \bibinfo {author} {\bibfnamefont {P.}~\bibnamefont {Bordone}},
  \bibinfo {author} {\bibfnamefont {R.}~\bibnamefont {Brunetti}}, \bibinfo
  {author} {\bibfnamefont {C.}~\bibnamefont {Jacoboni}}, \ and\ \bibinfo
  {author} {\bibfnamefont {S.}~\bibnamefont {Reggiani}},\ }\bibfield  {title}
  {\enquote {\bibinfo {title} {Quantum logic gates based on coherent electron
  transport in quantum wires},}\ }\href {\doibase 10.1103/PhysRevLett.84.5912}
  {\bibfield  {journal} {\bibinfo  {journal} {Phys. Rev. Lett.}\ }\textbf
  {\bibinfo {volume} {84}},\ \bibinfo {pages} {5912} (\bibinfo {year}
  {2000})}\BibitemShut {NoStop}%
\bibitem [{\citenamefont {Ionicioiu}\ \emph {et~al.}(2001)\citenamefont
  {Ionicioiu}, \citenamefont {Amaratunga},\ and\ \citenamefont
  {Udrea}}]{Ionicioiu01}%
  \BibitemOpen
  \bibfield  {author} {\bibinfo {author} {\bibfnamefont {R.}\ \bibnamefont
  {Ionicioiu}}, \bibinfo {author} {\bibfnamefont {G.}\ \bibnamefont
  {Amaratunga}}, \ and\ \bibinfo {author} {\bibfnamefont {F.}\ \bibnamefont
  {Udrea}},\ }\bibfield  {title} {\enquote {\bibinfo {title} {Quantum
  computation with ballistic electrons},}\ }\href {\doibase
  10.1142/S0217979201003521} {\bibfield  {journal} {\bibinfo  {journal} {Int.
  J. Mod. Phys. B}\ }\textbf {\bibinfo {volume} {15}},\ \bibinfo {pages}
  {125--133} (\bibinfo {year} {2001})}\BibitemShut {NoStop}%
\bibitem [{\citenamefont {F{\`e}ve}\ \emph {et~al.}(2007)\citenamefont
  {F{\`e}ve}, \citenamefont {Mah{\'e}}, \citenamefont {Berroir}, \citenamefont
  {Kontos}, \citenamefont {Plaçais}, \citenamefont {Glattli}, \citenamefont
  {Cavanna}, \citenamefont {Etienne},\ and\ \citenamefont {Jin}}]{Feve07}%
  \BibitemOpen
  \bibfield  {author} {\bibinfo {author} {\bibfnamefont {G.}~\bibnamefont
  {F{\`e}ve}}, \bibinfo {author} {\bibfnamefont {A.}~\bibnamefont {Mah{\'e}}},
  \bibinfo {author} {\bibfnamefont {J.-M.}\ \bibnamefont {Berroir}}, \bibinfo
  {author} {\bibfnamefont {T.}~\bibnamefont {Kontos}}, \bibinfo {author}
  {\bibfnamefont {B.}~\bibnamefont {Plaçais}}, \bibinfo {author}
  {\bibfnamefont {D.~C.}\ \bibnamefont {Glattli}}, \bibinfo {author}
  {\bibfnamefont {A.}~\bibnamefont {Cavanna}}, \bibinfo {author} {\bibfnamefont
  {B.}~\bibnamefont {Etienne}}, \ and\ \bibinfo {author} {\bibfnamefont
  {Y.}~\bibnamefont {Jin}},\ }\bibfield  {title} {\enquote {\bibinfo {title}
  {An on-demand coherent single-electron source},}\ }\href {\doibase
  10.1126/science.1141243} {\bibfield  {journal} {\bibinfo  {journal}
  {Science}\ }\textbf {\bibinfo {volume} {316}},\ \bibinfo {pages} {1169--1172}
  (\bibinfo {year} {2007})}\BibitemShut {NoStop}%
\bibitem [{\citenamefont {Feng}\ \emph {et~al.}(2012)\citenamefont {Feng},
  \citenamefont {Qian}, \citenamefont {Huang},\ and\ \citenamefont
  {Li}}]{Feng12}%
  \BibitemOpen
  \bibfield  {author} {\bibinfo {author} {\bibfnamefont {Ji}~\bibnamefont
  {Feng}}, \bibinfo {author} {\bibfnamefont {Xiaofeng}\ \bibnamefont {Qian}},
  \bibinfo {author} {\bibfnamefont {Cheng-Wei}\ \bibnamefont {Huang}}, \ and\
  \bibinfo {author} {\bibfnamefont {Ju}~\bibnamefont {Li}},\ }\bibfield
  {title} {\enquote {\bibinfo {title} {Strain-engineered artificial atom as a
  broad-spectrum solar energy funnel},}\ }\href@noop {} {\bibfield  {journal}
  {\bibinfo  {journal} {Nat. Photon.}\ }\textbf {\bibinfo {volume} {6}},\
  \bibinfo {pages} {866--872} (\bibinfo {year} {2012})}\BibitemShut {NoStop}%
\bibitem [{\citenamefont {Manzeli}\ \emph {et~al.}(2015)\citenamefont
  {Manzeli}, \citenamefont {Allain}, \citenamefont {Ghadimi},\ and\
  \citenamefont {Kis}}]{Manzeli15}%
  \BibitemOpen
  \bibfield  {author} {\bibinfo {author} {\bibfnamefont {S.}\ \bibnamefont
  {Manzeli}}, \bibinfo {author} {\bibfnamefont {A.}\ \bibnamefont
  {Allain}}, \bibinfo {author} {\bibfnamefont {A.}\ \bibnamefont
  {Ghadimi}}, \ and\ \bibinfo {author} {\bibfnamefont {A.}\ \bibnamefont
  {Kis}},\ }\bibfield  {title} {\enquote {\bibinfo {title} {Piezoresistivity
  and strain-induced band gap tuning in atomically thin MoS2},}\ }\href
  {\doibase 10.1021/acs.nanolett.5b01689} {\bibfield  {journal} {\bibinfo
  {journal} {Nano Lett.}\ }\textbf {\bibinfo {volume} {15}},\ \bibinfo
  {pages} {5330--5335} (\bibinfo {year} {2015})}\BibitemShut {NoStop}%
\bibitem [{\citenamefont {Rold{\'a}n}\ \emph {et~al.}(2015)\citenamefont
  {Rold{\'a}n}, \citenamefont {Castellanos-Gomez}, \citenamefont {Cappelluti},\
  and\ \citenamefont {Guinea}}]{Roldan15}%
  \BibitemOpen
  \bibfield  {author} {\bibinfo {author} {\bibfnamefont {R.}\ \bibnamefont
  {Rold{\'a}n}}, \bibinfo {author} {\bibfnamefont {A.}\ \bibnamefont
  {Castellanos-Gomez}}, \bibinfo {author} {\bibfnamefont {E.}\
  \bibnamefont {Cappelluti}}, \ and\ \bibinfo {author} {\bibfnamefont
  {F.}\ \bibnamefont {Guinea}},\ }\bibfield  {title} {\enquote {\bibinfo
  {title} {Strain engineering in semiconducting two-dimensional crystals},}\
  }\href@noop {} {\bibfield  {journal} {\bibinfo  {journal} {J. Phys. Condens. Matter}\ }\textbf {\bibinfo {volume} {27}},\ \bibinfo
  {pages} {313201} (\bibinfo {year} {2015})}\BibitemShut {NoStop}%
\bibitem [{\citenamefont {Kern}\ \emph {et~al.}(2016)\citenamefont {Kern},
  \citenamefont {Niehues}, \citenamefont {Tonndorf}, \citenamefont {Schmidt},
  \citenamefont {Wigger}, \citenamefont {Schneider}, \citenamefont {Stiehm},
  \citenamefont {Michaelis~de Vasconcellos}, \citenamefont {Reiter},
  \citenamefont {Kuhn},\ and\ \citenamefont {Bratschitsch}}]{Kern16}%
  \BibitemOpen
  \bibfield  {author} {\bibinfo {author} {\bibfnamefont {J.}\
  \bibnamefont {Kern}}, \bibinfo {author} {\bibfnamefont {I.}\ \bibnamefont
  {Niehues}}, \bibinfo {author} {\bibfnamefont {P.}\ \bibnamefont
  {Tonndorf}}, \bibinfo {author} {\bibfnamefont {R.}\ \bibnamefont
  {Schmidt}}, \bibinfo {author} {\bibfnamefont {D.}\ \bibnamefont
  {Wigger}}, \bibinfo {author} {\bibfnamefont {R.}\ \bibnamefont
  {Schneider}}, \bibinfo {author} {\bibfnamefont {T.}\ \bibnamefont
  {Stiehm}}, \bibinfo {author} {\bibfnamefont {S.}\ \bibnamefont
  {Michaelis~de Vasconcellos}}, \bibinfo {author} {\bibfnamefont {D.~E.}\
  \bibnamefont {Reiter}}, \bibinfo {author} {\bibfnamefont {T.}\
  \bibnamefont {Kuhn}}, \ and\ \bibinfo {author} {\bibfnamefont {R.}\
  \bibnamefont {Bratschitsch}},\ }\bibfield  {title} {\enquote {\bibinfo
  {title} {Nanoscale positioning of single-photon emitters in atomically thin
  WSe2},}\ }\href {\doibase 10.1002/adma.201600560} {\bibfield  {journal}
  {\bibinfo  {journal} {Adv. Mater.}\ }\textbf {\bibinfo {volume}
  {28}},\ \bibinfo {pages} {7101--7105} (\bibinfo {year} {2016})}\BibitemShut
  {NoStop}%
\bibitem [{\citenamefont {Rosati}\ \emph {et~al.}(2015)\citenamefont {Rosati},
  \citenamefont {Dolcini},\ and\ \citenamefont {Rossi}}]{Rosati15}%
  \BibitemOpen
  \bibfield  {author} {\bibinfo {author} {\bibfnamefont {R.}\ \bibnamefont
  {Rosati}}, \bibinfo {author} {\bibfnamefont {F.}\ \bibnamefont
  {Dolcini}}, \ and\ \bibinfo {author} {\bibfnamefont {F.}\ \bibnamefont
  {Rossi}},\ }\bibfield  {title} {\enquote {\bibinfo {title} {Dispersionless
  propagation of electron wavepackets in single-walled carbon nanotubes},}\
  }\href {\doibase http://dx.doi.org/10.1063/1.4922739} {\bibfield  {journal}
  {\bibinfo  {journal} {Appl. Phys. Lett.}\ }\textbf {\bibinfo {volume}
  {106}},\ \bibinfo {eid} {243101} (\bibinfo {year} {2015})}\BibitemShut
  {NoStop}%
\bibitem [{\citenamefont {Rosati}\ \emph {et~al.}(2015)\citenamefont {Rosati},
  \citenamefont {Dolcini},\ and\ \citenamefont {Rossi}}]{Rosati15b}%
  \BibitemOpen
  \bibfield  {author} {\bibinfo {author} {\bibfnamefont {R.}\ \bibnamefont
  {Rosati}}, \bibinfo {author} {\bibfnamefont {F.}\ \bibnamefont
  {Dolcini}}, \ and\ \bibinfo {author} {\bibfnamefont {F.}\ \bibnamefont
  {Rossi}},\ }\bibfield  {title} {\enquote {\bibinfo {title} {Electron-phonon
  coupling in metallic carbon nanotubes: Dispersionless electron propagation
  despite dissipation},}\ }\href {\doibase 10.1103/PhysRevB.92.235423}
  {\bibfield  {journal} {\bibinfo  {journal} {Phys. Rev. B}\ }\textbf {\bibinfo
  {volume} {92}},\ \bibinfo {pages} {235423} (\bibinfo {year}
  {2015})}\BibitemShut {NoStop}%
\bibitem [{\citenamefont {Brum}\ and\ \citenamefont {Bastard}(1986)}]{Brum86}%
  \BibitemOpen
  \bibfield  {author} {\bibinfo {author} {\bibfnamefont {J.~A.}\ \bibnamefont
  {Brum}}\ and\ \bibinfo {author} {\bibfnamefont {G.}~\bibnamefont {Bastard}},\
  }\bibfield  {title} {\enquote {\bibinfo {title} {Resonant carrier capture by
  semiconductor quantum wells},}\ }\href {\doibase 10.1103/PhysRevB.33.1420}
  {\bibfield  {journal} {\bibinfo  {journal} {Phys. Rev. B}\ }\textbf {\bibinfo
  {volume} {33}},\ \bibinfo {pages} {1420--1423} (\bibinfo {year}
  {1986})}\BibitemShut {NoStop}%
\bibitem [{\citenamefont {Kuhn}\ and\ \citenamefont {Mahler}(1989)}]{Kuhn89}%
  \BibitemOpen
  \bibfield  {author} {\bibinfo {author} {\bibfnamefont {T.}~\bibnamefont
  {Kuhn}}\ and\ \bibinfo {author} {\bibfnamefont {G.}~\bibnamefont {Mahler}},\
  }\bibfield  {title} {\enquote {\bibinfo {title} {Carrier capture in quantum
  wells and its importance for ambipolar transport},}\ }\href@noop {}
  {\bibfield  {journal} {\bibinfo  {journal} {Solid-Stole Electron.}\
  }\textbf {\bibinfo {volume} {32}},\ \bibinfo {pages} {1851--1855} (\bibinfo
  {year} {1989})}\BibitemShut {NoStop}%
\bibitem [{\citenamefont {Preisel}\ and\ \citenamefont
  {Mo/rk}(1994)}]{Preisel94}%
  \BibitemOpen
  \bibfield  {author} {\bibinfo {author} {\bibfnamefont {M.}\ \bibnamefont
  {Preisel}}\ and\ \bibinfo {author} {\bibfnamefont {J.}\ \bibnamefont
  {M{\o}rk}},\ }\bibfield  {title} {\enquote {\bibinfo {title} {Phonon-mediated
  carrier capture in quantum well lasers},}\ }\href@noop {} {\bibfield
  {journal} {\bibinfo  {journal} {J. Appl. Phys.}\ }\textbf
  {\bibinfo {volume} {76}},\ \bibinfo {eid} {1691} (\bibinfo {year} {1994})}\BibitemShut {NoStop}%
\bibitem [{\citenamefont {Reiter}\ \emph {et~al.}(2009)\citenamefont {Reiter},
  \citenamefont {Sherman}, \citenamefont {Najmaie},\ and\ \citenamefont
  {Sipe}}]{Reiter09}%
  \BibitemOpen
  \bibfield  {author} {\bibinfo {author} {\bibfnamefont {D.~E.}\ \bibnamefont
  {Reiter}}, \bibinfo {author} {\bibfnamefont {E.~Ya.}\ \bibnamefont
  {Sherman}}, \bibinfo {author} {\bibfnamefont {A.}~\bibnamefont {Najmaie}}, \
  and\ \bibinfo {author} {\bibfnamefont {J.~E.}\ \bibnamefont {Sipe}},\
  }\bibfield  {title} {\enquote {\bibinfo {title} {Coherent control of electron
  propagation and capture in semiconductor heterostructures},}\ }\href
  {http://stacks.iop.org/0295-5075/88/i=6/a=67005} {\bibfield  {journal}
  {\bibinfo  {journal} {EPL}\ }\textbf {\bibinfo {volume}
  {88}},\ \bibinfo {pages} {67005} (\bibinfo {year} {2009})}\BibitemShut
  {NoStop}%
\bibitem [{\citenamefont {Ferreira}\ and\ \citenamefont
  {Bastard}(1999)}]{Ferreira99}%
  \BibitemOpen
  \bibfield  {author} {\bibinfo {author} {\bibfnamefont {R.}~\bibnamefont
  {Ferreira}}\ and\ \bibinfo {author} {\bibfnamefont {G.}~\bibnamefont
  {Bastard}},\ }\bibfield  {title} {\enquote {\bibinfo {title} {Phonon-assisted
  capture and intradot auger relaxation in quantum dots},}\ }\href@noop {}
  {\bibfield  {journal} {\bibinfo  {journal} {Appl. Phys. Lett.}\
  }\textbf {\bibinfo {volume} {74}},\ \bibinfo {eid} {2818} (\bibinfo {year} {1999})}\BibitemShut
  {NoStop}%
\bibitem [{\citenamefont {Magnusdottir}\ \emph {et~al.}(2003)\citenamefont
  {Magnusdottir}, \citenamefont {Bischoff}, \citenamefont {Uskov},\ and\
  \citenamefont {M\o{}rk}}]{Magnusdottir03}%
  \BibitemOpen
  \bibfield  {author} {\bibinfo {author} {\bibfnamefont {I.}~\bibnamefont
  {Magnusdottir}}, \bibinfo {author} {\bibfnamefont {S.}~\bibnamefont
  {Bischoff}}, \bibinfo {author} {\bibfnamefont {A.~V.}\ \bibnamefont {Uskov}},
  \ and\ \bibinfo {author} {\bibfnamefont {J.}~\bibnamefont {M\o{}rk}},\
  }\bibfield  {title} {\enquote {\bibinfo {title} {Geometry dependence of Auger
  carrier capture rates into cone-shaped self-assembled quantum dots},}\ }\href
  {\doibase 10.1103/PhysRevB.67.205326} {\bibfield  {journal} {\bibinfo
  {journal} {Phys. Rev. B}\ }\textbf {\bibinfo {volume} {67}},\ \bibinfo
  {pages} {205326} (\bibinfo {year} {2003})}\BibitemShut {NoStop}%
\bibitem [{\citenamefont {Nielsen}\ \emph {et~al.}(2004)\citenamefont
  {Nielsen}, \citenamefont {Gartner},\ and\ \citenamefont
  {Jahnke}}]{Nielsen04}%
  \BibitemOpen
  \bibfield  {author} {\bibinfo {author} {\bibfnamefont {T.~R.}\ \bibnamefont
  {Nielsen}}, \bibinfo {author} {\bibfnamefont {P.}~\bibnamefont {Gartner}}, \
  and\ \bibinfo {author} {\bibfnamefont {F.}~\bibnamefont {Jahnke}},\
  }\bibfield  {title} {\enquote {\bibinfo {title} {Many-body theory of carrier
  capture and relaxation in semiconductor quantum-dot lasers},}\ }\href
  {\doibase 10.1103/PhysRevB.69.235314} {\bibfield  {journal} {\bibinfo
  {journal} {Phys. Rev. B}\ }\textbf {\bibinfo {volume} {69}},\ \bibinfo
  {pages} {235314} (\bibinfo {year} {2004})}\BibitemShut {NoStop}%
\bibitem [{\citenamefont {Herbst}\ \emph {et~al.}(2003)\citenamefont {Herbst},
  \citenamefont {Glanemann}, \citenamefont {Axt},\ and\ \citenamefont
  {Kuhn}}]{Herbst03}%
  \BibitemOpen
  \bibfield  {author} {\bibinfo {author} {\bibfnamefont {M.}~\bibnamefont
  {Herbst}}, \bibinfo {author} {\bibfnamefont {M.}~\bibnamefont {Glanemann}},
  \bibinfo {author} {\bibfnamefont {V.~M.}\ \bibnamefont {Axt}}, \ and\
  \bibinfo {author} {\bibfnamefont {T.}~\bibnamefont {Kuhn}},\ }\bibfield
  {title} {\enquote {\bibinfo {title} {Electron-phonon quantum kinetics for
  spatially inhomogeneous excitations},}\ }\href {\doibase
  10.1103/PhysRevB.67.195305} {\bibfield  {journal} {\bibinfo  {journal} {Phys.
  Rev. B}\ }\textbf {\bibinfo {volume} {67}},\ \bibinfo {pages} {195305}
  (\bibinfo {year} {2003})}\BibitemShut {NoStop}%
\bibitem [{\citenamefont {Steininger}\ \emph {et~al.}(1996)\citenamefont
  {Steininger}, \citenamefont {Knorr}, \citenamefont {Stroucken}, \citenamefont
  {Thomas},\ and\ \citenamefont {Koch}}]{Steininger96}%
  \BibitemOpen
  \bibfield  {author} {\bibinfo {author} {\bibfnamefont {F.}~\bibnamefont
  {Steininger}}, \bibinfo {author} {\bibfnamefont {A.}~\bibnamefont {Knorr}},
  \bibinfo {author} {\bibfnamefont {T.}~\bibnamefont {Stroucken}}, \bibinfo
  {author} {\bibfnamefont {P.}~\bibnamefont {Thomas}}, \ and\ \bibinfo {author}
  {\bibfnamefont {S.~W.}\ \bibnamefont {Koch}},\ }\bibfield  {title} {\enquote
  {\bibinfo {title} {Dynamic evolution of spatiotemporally localized electronic
  wave packets in semiconductor quantum wells},}\ }\href {\doibase
  10.1103/PhysRevLett.77.550} {\bibfield  {journal} {\bibinfo  {journal} {Phys.
  Rev. Lett.}\ }\textbf {\bibinfo {volume} {77}},\ \bibinfo {pages} {550--553}
  (\bibinfo {year} {1996})}\BibitemShut {NoStop}%
\bibitem [{\citenamefont {Reiter}\ \emph {et~al.}(2007)\citenamefont {Reiter},
  \citenamefont {Glanemann}, \citenamefont {Axt},\ and\ \citenamefont
  {Kuhn}}]{Reiter07a}%
  \BibitemOpen
  \bibfield  {author} {\bibinfo {author} {\bibfnamefont {D.}~\bibnamefont
  {Reiter}}, \bibinfo {author} {\bibfnamefont {M.}~\bibnamefont {Glanemann}},
  \bibinfo {author} {\bibfnamefont {V.~M.}\ \bibnamefont {Axt}}, \ and\
  \bibinfo {author} {\bibfnamefont {T.}~\bibnamefont {Kuhn}},\ }\bibfield
  {title} {\enquote {\bibinfo {title} {Spatiotemporal dynamics in optically
  excited quantum wire-dot systems: Capture, escape, and wave-front
  dynamics},}\ }\href {\doibase 10.1103/PhysRevB.75.205327} {\bibfield
  {journal} {\bibinfo  {journal} {Phys. Rev. B}\ }\textbf {\bibinfo {volume}
  {75}},\ \bibinfo {pages} {205327} (\bibinfo {year} {2007})}\BibitemShut
  {NoStop}%
\bibitem [{\citenamefont {Rosati}\ \emph {et~al.}(2014)\citenamefont {Rosati},
  \citenamefont {Iotti}, \citenamefont {Dolcini},\ and\ \citenamefont
  {Rossi}}]{Rosati14e}%
  \BibitemOpen
  \bibfield  {author} {\bibinfo {author} {\bibfnamefont {R.}\ \bibnamefont
  {Rosati}}, \bibinfo {author} {\bibfnamefont {R.~C.}\ \bibnamefont
  {Iotti}}, \bibinfo {author} {\bibfnamefont {F.}\ \bibnamefont
  {Dolcini}}, \ and\ \bibinfo {author} {\bibfnamefont {F.}\ \bibnamefont
  {Rossi}},\ }\bibfield  {title} {\enquote {\bibinfo {title} {Derivation of
  nonlinear single-particle equations via many-body lindblad superoperators: A
  density-matrix approach},}\ }\href {\doibase 10.1103/PhysRevB.90.125140}
  {\bibfield  {journal} {\bibinfo  {journal} {Phys. Rev. B}\ }\textbf {\bibinfo
  {volume} {90}},\ \bibinfo {pages} {125140} (\bibinfo {year}
  {2014})}\BibitemShut {NoStop}%
\bibitem [{\citenamefont {Taj}\ \emph {et~al.}(2009)\citenamefont {Taj},
  \citenamefont {Iotti},\ and\ \citenamefont {Rossi}}]{Taj09b}%
  \BibitemOpen
  \bibfield  {author} {\bibinfo {author} {\bibfnamefont {D.}~\bibnamefont
  {Taj}}, \bibinfo {author} {\bibfnamefont {R.~C.}\ \bibnamefont {Iotti}}, \
  and\ \bibinfo {author} {\bibfnamefont {F.}~\bibnamefont {Rossi}},\ }\bibfield
   {title} {\enquote {\bibinfo {title} {Microscopic modeling of energy
  relaxation and decoherence in quantum optoelectronic devices at the
  nanoscale},}\ }\href {\doibase 10.1140/epjb/e2009-00363-4} {\bibfield
  {journal} {\bibinfo  {journal} {Eur. Phys. J. B}\ }\textbf {\bibinfo {volume}
  {72}},\ \bibinfo {pages} {305--322} (\bibinfo {year} {2009})}\BibitemShut
  {NoStop}%
\bibitem [{\citenamefont {Rossi}(2011)}]{b-Rossi11}%
  \BibitemOpen
  \bibfield  {author} {\bibinfo {author} {\bibfnamefont {F.}~\bibnamefont
  {Rossi}},\ }\href@noop {} {\emph {\bibinfo {title} {Theory of Semiconductor
  Quantum Devices: Microscopic Modeling and Simulation Strategies}}}\ (\bibinfo
   {publisher} {Springer, Berlin},\ \bibinfo {year} {2011})\BibitemShut {NoStop}%
\bibitem [{\citenamefont {Rosati}\ and\ \citenamefont
  {Rossi}(2013)}]{Rosati13b}%
  \BibitemOpen
  \bibfield  {author} {\bibinfo {author} {\bibfnamefont {R.}\ \bibnamefont
  {Rosati}}\ and\ \bibinfo {author} {\bibfnamefont {F.}\ \bibnamefont
  {Rossi}},\ }\bibfield  {title} {\enquote {\bibinfo {title} {Microscopic
  modeling of scattering quantum non-locality in semiconductor
  nanostructures},}\ }\href {\doibase 10.1063/1.4821158} {\bibfield  {journal}
  {\bibinfo  {journal} {Appl. Phys. Lett.}\ }\textbf {\bibinfo {volume}
  {103}},\ \bibinfo {pages} {113105} (\bibinfo {year} {2013})}\BibitemShut
  {NoStop}%
\bibitem [{\citenamefont {Rosati}\ and\ \citenamefont
  {Rossi}(2014{\natexlab{a}})}]{Rosati14a}%
  \BibitemOpen
  \bibfield  {author} {\bibinfo {author} {\bibfnamefont {R.}\ \bibnamefont
  {Rosati}}\ and\ \bibinfo {author} {\bibfnamefont {F.}\ \bibnamefont
  {Rossi}},\ }\bibfield  {title} {\enquote {\bibinfo {title} {Quantum diffusion
  due to scattering non-locality in nanoscale semiconductors},}\ }\href
  {\doibase 10.1209/0295-5075/105/17010} {\bibfield  {journal} {\bibinfo
  {journal} {EPL}\ }\textbf {\bibinfo {volume} {105}},\ \bibinfo {pages}
  {17010} (\bibinfo {year} {2014}{\natexlab{a}})}\BibitemShut {NoStop}%
\bibitem [{\citenamefont {Rosati}\ and\ \citenamefont
  {Rossi}(2014)}]{Rosati14b}%
  \BibitemOpen
  \bibfield  {author} {\bibinfo {author} {\bibfnamefont {R.}\ \bibnamefont
  {Rosati}}\ and\ \bibinfo {author} {\bibfnamefont {F.}\ \bibnamefont
  {Rossi}},\ }\bibfield  {title} {\enquote {\bibinfo {title} {Scattering
  nonlocality in quantum charge transport: Application to semiconductor
  nanostructures},}\ }\href {\doibase 10.1103/PhysRevB.89.205415} {\bibfield
  {journal} {\bibinfo  {journal} {Phys. Rev. B}\ }\textbf {\bibinfo {volume}
  {89}},\ \bibinfo {pages} {205415} (\bibinfo {year} {2014})}\BibitemShut
  {NoStop}%
\bibitem [{\citenamefont {Dolcini}\ \emph {et~al.}(2013)\citenamefont
  {Dolcini}, \citenamefont {Iotti},\ and\ \citenamefont {Rossi}}]{Dolcini13a}%
  \BibitemOpen
  \bibfield  {author} {\bibinfo {author} {\bibfnamefont {F.}\
  \bibnamefont {Dolcini}}, \bibinfo {author} {\bibfnamefont {R.~C.}\
  \bibnamefont {Iotti}}, \ and\ \bibinfo {author} {\bibfnamefont {F.}\
  \bibnamefont {Rossi}},\ }\bibfield  {title} {\enquote {\bibinfo {title}
  {Interplay between energy dissipation and reservoir-induced thermalization in
  nonequilibrium quantum nanodevices},}\ }\href {\doibase
  10.1103/PhysRevB.88.115421} {\bibfield  {journal} {\bibinfo  {journal} {Phys.
  Rev. B}\ }\textbf {\bibinfo {volume} {88}},\ \bibinfo {pages} {115421}
  (\bibinfo {year} {2013})}\BibitemShut {NoStop}%
\bibitem [{\citenamefont {Glanemann}\ \emph {et~al.}(2004)\citenamefont
  {Glanemann}, \citenamefont {Axt},\ and\ \citenamefont {Kuhn}}]{Glanemann04}%
  \BibitemOpen
  \bibfield  {author} {\bibinfo {author} {\bibfnamefont {M.}~\bibnamefont
  {Glanemann}}, \bibinfo {author} {\bibfnamefont {V.~M.}\ \bibnamefont {Axt}}, \
  and\ \bibinfo {author} {\bibfnamefont {T.}~\bibnamefont {Kuhn}},\ }\bibfield
  {title} {\enquote {\bibinfo {title} {Thermal escape and capture processes in
  quantum wire-dot structures},}\ }\href
  {http://stacks.iop.org/0268-1242/19/i=4/a=077} {\bibfield  {journal}
  {\bibinfo  {journal} {Semicond. Sci. Technol.}\ }\textbf
  {\bibinfo {volume} {19}},\ \bibinfo {pages} {S229} (\bibinfo {year}
  {2004})}\BibitemShut {NoStop}%
\bibitem [{\citenamefont {Raimond}\ \emph {et~al.}(2001)\citenamefont
  {Raimond}, \citenamefont {Brune},\ and\ \citenamefont {Haroche}}]{Raimond01}%
  \BibitemOpen
  \bibfield  {author} {\bibinfo {author} {\bibfnamefont {J.~M.}\ \bibnamefont
  {Raimond}}, \bibinfo {author} {\bibfnamefont {M.}~\bibnamefont {Brune}}, \
  and\ \bibinfo {author} {\bibfnamefont {S.}~\bibnamefont {Haroche}},\
  }\bibfield  {title} {\enquote {\bibinfo {title} {Manipulating quantum
  entanglement with atoms and photons in a cavity},}\ }\href {\doibase
  10.1103/RevModPhys.73.565} {\bibfield  {journal} {\bibinfo  {journal} {Rev.
  Mod. Phys.}\ }\textbf {\bibinfo {volume} {73}},\ \bibinfo {pages} {565--582}
  (\bibinfo {year} {2001})}\BibitemShut {NoStop}%
\bibitem [{\citenamefont {Verzelen}\ \emph {et~al.}(2002)\citenamefont
  {Verzelen}, \citenamefont {Ferreira},\ and\ \citenamefont
  {Bastard}}]{Verzelen02}%
  \BibitemOpen
  \bibfield  {author} {\bibinfo {author} {\bibfnamefont {O.}~\bibnamefont
  {Verzelen}}, \bibinfo {author} {\bibfnamefont {R.}~\bibnamefont {Ferreira}},
  \ and\ \bibinfo {author} {\bibfnamefont {G.}~\bibnamefont {Bastard}},\
  }\bibfield  {title} {\enquote {\bibinfo {title} {Excitonic polarons in
  semiconductor quantum dots},}\ }\href {\doibase
  10.1103/PhysRevLett.88.146803} {\bibfield  {journal} {\bibinfo  {journal}
  {Phys. Rev. Lett.}\ }\textbf {\bibinfo {volume} {88}},\ \bibinfo {pages}
  {146803} (\bibinfo {year} {2002})}\BibitemShut {NoStop}%
\bibitem [{\citenamefont {Hameau}\ \emph {et~al.}(1999)\citenamefont {Hameau},
  \citenamefont {Guldner}, \citenamefont {Verzelen}, \citenamefont {Ferreira},
  \citenamefont {Bastard}, \citenamefont {Zeman}, \citenamefont
  {Lema\^{\i}tre},\ and\ \citenamefont {G\'erard}}]{Hameau99}%
  \BibitemOpen
  \bibfield  {author} {\bibinfo {author} {\bibfnamefont {S.}~\bibnamefont
  {Hameau}}, \bibinfo {author} {\bibfnamefont {Y.}~\bibnamefont {Guldner}},
  \bibinfo {author} {\bibfnamefont {O.}~\bibnamefont {Verzelen}}, \bibinfo
  {author} {\bibfnamefont {R.}~\bibnamefont {Ferreira}}, \bibinfo {author}
  {\bibfnamefont {G.}~\bibnamefont {Bastard}}, \bibinfo {author} {\bibfnamefont
  {J.}~\bibnamefont {Zeman}}, \bibinfo {author} {\bibfnamefont
  {A.}~\bibnamefont {Lema\^{\i}tre}}, \ and\ \bibinfo {author} {\bibfnamefont
  {J.~M.}\ \bibnamefont {G\'erard}},\ }\bibfield  {title} {\enquote {\bibinfo
  {title} {Strong electron-phonon coupling regime in quantum dots: Evidence for
  everlasting resonant polarons},}\ }\href {\doibase
  10.1103/PhysRevLett.83.4152} {\bibfield  {journal} {\bibinfo  {journal}
  {Phys. Rev. Lett.}\ }\textbf {\bibinfo {volume} {83}},\ \bibinfo {pages}
  {4152--4155} (\bibinfo {year} {1999})}\BibitemShut {NoStop}%
\bibitem [{\citenamefont {Badalyan}\ and\ \citenamefont
  {Levinson}(1988)}]{Badalyan88}%
  \BibitemOpen
  \bibfield  {author} {\bibinfo {author} {\bibfnamefont {S.~M.}~\bibnamefont
  {Badalyan}}\ and\ \bibinfo {author} {\bibfnamefont {Y.~B.}~\bibnamefont
  {Levinson}},\ }\bibfield  {title} {\enquote {\bibinfo {title} {Bound states
  of electron and optical phonon in a quantum well},}\ }\href@noop {}
  {\bibfield  {journal} {\bibinfo  {journal} {Zh. Eksp. Teor. Fiz}\ }\textbf
  {\bibinfo {volume} {94}},\ \bibinfo {pages} {378} (\bibinfo {year}
  {1988})}\BibitemShut {NoStop}%
\bibitem [{\citenamefont {Krummheuer}\ \emph {et~al.}(2002)\citenamefont
  {Krummheuer}, \citenamefont {Axt},\ and\ \citenamefont
  {Kuhn}}]{Krummheuer02}%
  \BibitemOpen
  \bibfield  {author} {\bibinfo {author} {\bibfnamefont {B.}~\bibnamefont
  {Krummheuer}}, \bibinfo {author} {\bibfnamefont {V.~M.}\ \bibnamefont {Axt}},
  \ and\ \bibinfo {author} {\bibfnamefont {T.}~\bibnamefont {Kuhn}},\
  }\bibfield  {title} {\enquote {\bibinfo {title} {Theory of pure dephasing and
  the resulting absorption line shape in semiconductor quantum dots},}\ }\href
  {\doibase 10.1103/PhysRevB.65.195313} {\bibfield  {journal} {\bibinfo
  {journal} {Phys. Rev. B}\ }\textbf {\bibinfo {volume} {65}},\ \bibinfo
  {pages} {195313} (\bibinfo {year} {2002})}\BibitemShut {NoStop}%
\bibitem [{\citenamefont {Zimmermann}\ and\ \citenamefont
  {Wauer}(1994)}]{Zimmermann94}%
  \BibitemOpen
  \bibfield  {author} {\bibinfo {author} {\bibfnamefont {R.}~\bibnamefont
  {Zimmermann}}\ and\ \bibinfo {author} {\bibfnamefont {J.}~\bibnamefont
  {Wauer}},\ }\bibfield  {title} {\enquote {\bibinfo {title} {Non-Markovian
  relaxation in semiconductors: An exactly soluble model},}\ }\href {\doibase
  http://dx.doi.org/10.1016/0022-2313(94)90413-8} {\bibfield  {journal}
  {\bibinfo  {journal} {JOL}\ }\textbf {\bibinfo {volume}
  {58}},\ \bibinfo {pages} {271 -- 274} (\bibinfo {year} {1994})}\BibitemShut
  {NoStop}%
\bibitem [{\citenamefont {Kr{\"u}gel}\ \emph {et~al.}(2006)\citenamefont
  {Kr{\"u}gel}, \citenamefont {Axt},\ and\ \citenamefont {Kuhn}}]{Krugel06}%
  \BibitemOpen
  \bibfield  {author} {\bibinfo {author} {\bibfnamefont {A.}~\bibnamefont
  {Kr{\"u}gel}}, \bibinfo {author} {\bibfnamefont {V.~M.}\ \bibnamefont {Axt}},
  \ and\ \bibinfo {author} {\bibfnamefont {T.}~\bibnamefont {Kuhn}},\
  }\bibfield  {title} {\enquote {\bibinfo {title} {Back action of
  nonequilibrium phonons on the optically induced dynamics in semiconductor
  quantum dots},}\ }\href@noop {} {\bibfield  {journal} {\bibinfo  {journal}
  {Phys. Rev. B}\ }\textbf {\bibinfo {volume} {73}},\ \bibinfo {pages} {035302}
  (\bibinfo {year} {2006})}\BibitemShut {NoStop}%
\bibitem [{\citenamefont {Magnusdottir}\ \emph {et~al.}(2002)\citenamefont
  {Magnusdottir}, \citenamefont {Uskov}, \citenamefont {Ferreira},
  \citenamefont {Bastard}, \citenamefont {Mørk},\ and\ \citenamefont
  {Tromborg}}]{Magnusdottir02}%
  \BibitemOpen
  \bibfield  {author} {\bibinfo {author} {\bibfnamefont {I.}~\bibnamefont
  {Magnusdottir}}, \bibinfo {author} {\bibfnamefont {A.~V.}\ \bibnamefont
  {Uskov}}, \bibinfo {author} {\bibfnamefont {R.}~\bibnamefont {Ferreira}},
  \bibinfo {author} {\bibfnamefont {G.}~\bibnamefont {Bastard}}, \bibinfo
  {author} {\bibfnamefont {J.}~\bibnamefont {M{\o}rk}}, \ and\ \bibinfo {author}
  {\bibfnamefont {B.}~\bibnamefont {Tromborg}},\ }\bibfield  {title} {\enquote
  {\bibinfo {title} {Influence of quasibound states on the carrier capture in
  quantum dots},}\ }\href@noop {} {\bibfield  {journal} {\bibinfo  {journal}
  {Appl. Phys. Lett. }\ }\textbf {\bibinfo {volume} {81}},\ \bibinfo {eid} {4318} (\bibinfo {year}
  {2002})}\BibitemShut {NoStop}%
\end{thebibliography}

%

\end{document}